%% file: sirius_ngc6752.tex
\documentclass[twocolumn]{aastex63}

\usepackage{multirow}

\received{}
\revised{}
\accepted{}
\submitjournal{ApJ}

\shorttitle{Self-consistent analysis of stellar clusters: the halo globular cluster NGC 6752}
\shortauthors{Souza et al.}
\graphicspath{{./}{figs/}}
\graphicspath{{./}{figs/ngc6752/}}
\graphicspath{{./}{figs/synthetic50/}}

\begin{document}

\title{Self-consistent analysis of stellar clusters: An application  to {\it HST} data of the halo globular cluster NGC~6752}

\correspondingauthor{Stefano O. Souza}
\email{stefano.souza@usp.br}

\author[0000-0001-8052-969X]{S.~O. Souza}
\affil{Universidade de S\~ao Paulo, IAG, Rua do Mat\~ao 1226, Cidade
Universit\'aria, S\~ao Paulo 05508-900, Brazil}

\author[0000-0002-7435-8748]{L.~O. Kerber}
\affil{Universidade de S\~ao Paulo, IAG, Rua do Mat\~ao 1226, Cidade
Universit\'aria, S\~ao Paulo 05508-900, Brazil}
\affil{Universidade Estadual de Santa Cruz, Rodovia Jorge Amado km 16,
Ilh\'eus 45662-000, Brazil}

\author[0000-0001-9264-4417]{B. Barbuy} 
\affil{Universidade de S\~ao Paulo, IAG, Rua do Mat\~ao 1226, Cidade
Universit\'aria, S\~ao Paulo 05508-900, Brazil}

\author[0000-0002-5974-3998]{A. P\'erez-Villegas} 
\affil{Universidade de S\~ao Paulo, IAG, Rua do Mat\~ao 1226, Cidade
Universit\'aria, S\~ao Paulo 05508-900, Brazil}

\author[0000-0002-4778-9243]{R.~A.~P. Oliveira}
\affiliation{Universidade de S\~ao Paulo, IAG, Rua do Mat\~ao 1226, Cidade
Universit\'aria, S\~ao Paulo 05508-900, Brazil}

\author[0000-0003-1149-3659]{D. Nardiello}
\affil{Dipartimento di Fisica e Astronomia `Galileo Galilei', Universit\`a
di Padova, Vicolo dell'Osservatorio 3, Padova, I-35122, Italy}
\affil{Istituto Nazionale di Astrofisica\,--\,Osservatorio Astronomico di
Padova, Vicolo dell'Osservatorio 5, Padova, I-35122, Italy}
\affil{Aix Marseille Universit\'e, CNRS, CNES, LAM, Marseille, France}



\begin{abstract}


The Bayesian isochrone fitting using the Markov chain Monte Carlo algorithm is applied, to derive the probability distribution of the parameters age, metallicity, reddening, and absolute distance modulus. We introduce the \texttt{SIRIUS} code by means of simulated color-magnitude diagrams, including the analysis of  multiple stellar populations. The population tagging is applied from the red giant branch to the bottom of the main sequence. Through sanity checks using synthetic {\it HST} color-magnitude diagrams of globular clusters we verify the code reliability in the context of simple and multiple stellar populations. In such tests, the formal uncertainties in age or age difference, metallicity, reddening, and absolute distance modulus can reach $400$ Myr, $0.03$ dex, $0.01$ mag, and $0.03$ mag, respectively. We apply the method to analyse NGC~6752, using Dartmouth stellar evolutionary models. Assuming a single stellar population, we derive an age of $13.7\pm0.5$ Gyr and a distance of $d_{\odot}=4.11\pm 0.08$ kpc, with the latter in agreement within $~3\sigma$ with the inverse Gaia parallax. In the analysis of the multiple stellar populations, three {populations} are clearly identified. From the Chromosome Map and UV/Optical two-color diagrams inspection, we found a fraction of stars of $25\pm5$, $46\pm7$, and $29\pm5$ per cent, for the first, second, and third generations, respectively. These fractions are in good agreement with the literature. An age difference of $500\pm410$ Myr between the first and the third generation is found, with the uncertainty decreasing to $400$ Myr when the helium enhancement is taken into account.
\end{abstract}


\section{Introduction}



The study of stellar clusters has implications in a wide variety of astrophysical topics, which includes star formation, stellar evolution and nucleosynthesis, stellar dynamics, Galactic structure, and galaxy formation and evolution.%
\citep[e.g.][]{vandenberg13,barbuy18}. 

With the advent of space-based telescopes, in particular the {\it Hubble Space Telescope} ({\it HST}) and more recently the {\it Gaia} Data Release 2 \citep[DR2,][]{gaia18a}, as well as multi-object and high-resolution spectrographs, a wealth of high-quality and spatially resolved data have been collected for Milky Way globular and open clusters (GCs and OCs), and for stellar clusters in neighbouring galaxies. Combined with sophisticated analysis, these data have opened an unprecedented opportunity for very accurate physical parameter derivation. 

Milky Way globular clusters (GCs) formed during the early stages of the Galaxy formation \citep[e.g.][]{vandenberg13,barbuy18}
are studied in the present work.


The phenomenon of multiple stellar populations (MPs) was observed for the first time by \citet{osborn71} from 
CN-band strengths, but at the time this was not identified as due to the presence of two
stellar populations. Later,  MPs were clearly revealed by \citep[eg. ][]{lee99,bedin04,piotto05,milone17}, and
hints on self-enrichment to explain abundance variations within a GC were discussed by \citet{gratton04}. 
Evidence of MPs from spectroscopic work was reviewed by \citet[][and references therein]{carretta19}.
The photometric counterpart of the CN anomaly is detectable in the ultraviolet (UV) filters \citep{piotto15, lee19}. These filters are sensitive to C, N, and O abundances, allowing to disentangle the different stellar populations \citep{piotto15}.

With the purpose of correlating the cluster age with the presence of MPs, \citet{martocchia18,martocchia19} analyzed a sample of Magellanic Clouds (MCs) and MW clusters. They estimated the N abundance spread in CMDs, which is an indicator of the presence of MPs, and found that clusters older than $\sim 2$ Gyr host MPs, while those younger than this age show no evidence of spread in N abundance. On the other hand, it is known that the presence of MPs is related to the mass of the cluster \citep[][]{milone17}. For this reason, age cannot be the only parameter to constrain the presence of MPs. This fact is evident for the case of Berkeley 39 \citep[][]{martocchia18} and Lindsay 38 \citep[][]{martocchia19}, both having an age of $\sim 6.5$ Gyr, without showing N abundance spread. Another counterexample was given by \citet{lagioia19}, having  found that the GC Terzan 7 is consistent with a single stellar population (SSP), despite a relatively old age and high mass. Therefore, the study of MPs helps understanding the formation and evolution of stellar systems in general.




Isochrone fitting to CMDs has been extensively used to obtain the star cluster properties age, distance modulus, and reddening. Previously, a visual method known as ``chi-by-eye'' was usually employed to fit theoretical isochrones to CMDs. Later on, to benefit from improved data quality and to extract physical parameters with meaningful uncertainties, several statistical isochrone fitting techniques were developed, most of them based on $\chi^{2}$, maximum likelihood statistics, or Bayesian approach \citep[][]{kerber05,naylor06,vonHippel06,hernandez08, monteiro10}. In almost all these developments, synthetic CMDs are employed for validation of the methods. 

The Bayesian approach has the advantage of being able to get distributions and to explore the information \textit{a priori}
about the data or models. Recent examples of isochrone fitting codes using Bayesian inference are \texttt{ASteCA} \citep{perren15} and \texttt{BASE-9} 
\citep{stenning16}, where the latter allows analysis of MPs to derive their difference on the helium content (Y). \cite{ramirez-siordia19} also applied the Bayes' theorem to a Monte Carlo method to get the posterior distributions of the same parameters as \texttt{BASE-9}, neglecting helium enhancements. They applied their software to the scarce stellar populations of ultra-faint dwarf galaxies and LMC star clusters.


In the present work, we carry out a detailed analysis of CMDs assuming both cases 
of clusters as SSPs and MPs. With this purpose, 
we developed the code named \texttt{SIRIUS}\footnote{The code is available upon request to the authors.}, standing for {\bf S}tatistical {\bf I}nference of physical pa{\bf R}ameters of s{\bf I}ngle and m{\bf U}ltiple populations in {\bf S}tellar clusters, to extract information on a stellar cluster from its CMDs. 
The \texttt{SIRIUS} code was applied to analyse NGC~6752, with data from the \textit{HST} UV Legacy Survey of Galactic GCs  \citep{piotto15}.  \citep{gratton03} obtained for this halo GC an age of $13.4 \pm 1.1$  and  
Carretta et al. (2012) found three distinct stellar populations \citep{milone13}.
Whereas the precision in parameter derivation from CMDs has been improving, it is also important to stress that a new era is now open: the age difference between stellar populations in a GC can give us a better understanding on its formation. 

This work is organized as follows. In Section \ref{sec:sirius} the SIRIUS code is described in detail. Experiments to check the validity of the method and analysis of sources of uncertainties are presented in Section \ref{sec:tests}. An application to  {\it HST} data of the halo GC NGC~6752 is presented in Section \ref{sec:app}. Conclusions are drawn in Section \ref{sec:conclusion}.

\section{The \texttt{SIRIUS} code}\label{sec:sirius}

\begin{figure*}
    \centering
    \includegraphics[scale=0.54]{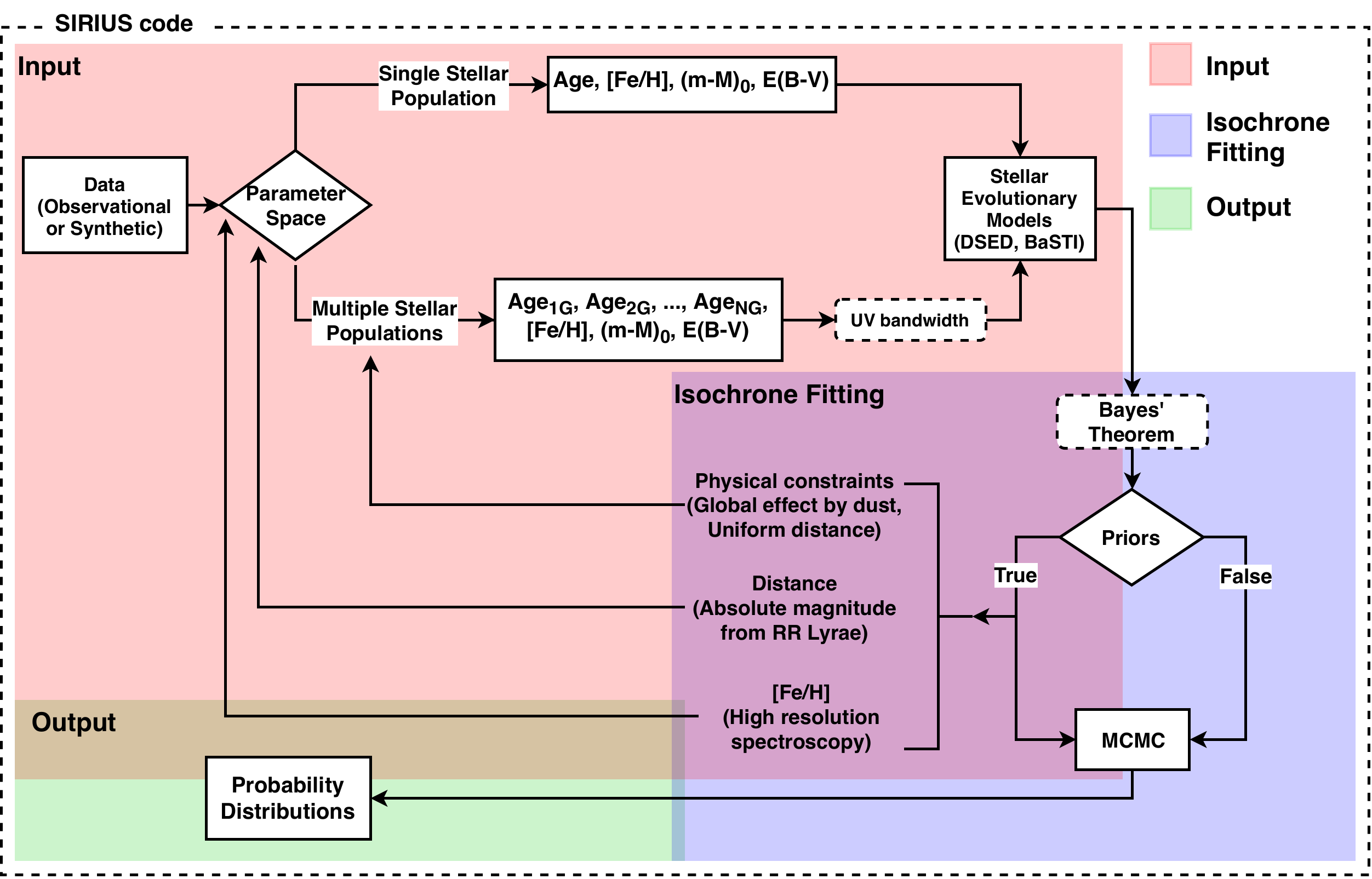}
    \caption{\texttt{SIRIUS} flow-chart shows the steps to perform the isochrone fitting. }
    \label{fig:workflow}
\end{figure*}

This section gives a detailed description of the \texttt{SIRIUS} code, built to carry out isochrone fitting to CMDs, following the flow-chart presented in Figure~\ref{fig:workflow}.

\subsection{Color-Magnitude Diagram Data}\label{subsec:Data}

\texttt{SIRIUS} was designed to analyse stellar clusters, applied here both to synthetic data and to
observed data. \texttt{SIRIUS} has already been successfully applied to derive the parameters of two bulge GCs. For HP\,1,  a multi-band ($K_{\rm S}$ and $J$ from Gemini-GSAOI+GeMS, and F606W from {\it HST}-ACS) isochrone fitting was applied \citep{kerber19}. For ESO 456-SC38, {\it HST} photometry in the filters F606W from ACS and F110W from WFC3, and FORS2@VLT photometry in V and I were used \citep{ortolani19}. These studies confirmed that HP~1 and
ESO~456-SC38 are among the oldest GCs in the Milky Way, with an age of $\sim12.8$ Gyr.

\texttt{SIRIUS} can create synthetic CMDs using the following method.
 The Monte Carlo algorithm  is used to generate random data from a given probability distribution, and can be applied to describe many physical systems. In the case of CMDs of stellar clusters the main probability distribution of the system is the initial mass function (IMF), here adopted to be the Kroupa IMF \citep{kroupa01}. The method to generate a sample of data similar to a stellar cluster is called as Synthetic CMD \citep{kerber07}. Points are randomly generated and interpolated in mass within theoretical points of isochrones. From an error function, these random points are dispersed by Gaussian distributions to simulate the spread seen in observed CMDs.

\subsection{Stellar evolutionary models and Parameter space}

The library of isochrones adopted include two sets of stellar evolutionary models: DSED\footnote
{\url{http://stellar.dartmouthThe.edu/models/grid.html}} \citep[Dartmouth Stellar Evolutionary Database -][]{dotter08} and
BaSTI\footnote
{\url{http://basti.oa-teramo.inaf.it/}} \citep[A Bag of Stellar Tracks and Isochrones -][]{pietrinferni06}. We perform linear regressions to interpolate the isochrones in steps of $0.1$ Gyr in age in the range of $10.0$ to $15.0$ Gyr, and $0.01$ dex in [Fe/H] in the range of $-2.00$ $<$ [Fe/H] $<$ $0.00$\footnote{The usual notation [Fe/H]=log(Fe/H)$_{star}$-log(Fe/H)$_{\odot}$ is adopted.}. 
It is relevant to mention that the range and step size of age we adopted here are consistent with the context of Galactic GCs. For the case of younger stellar clusters, e.g. MC clusters, the age range should allow ages below $10$ Gyr, and the step size should be narrower than the value used here.


The simple $\chi^2$ isochrone fitting procedures do not necessarily represent a physical interpretation of a GC CMD. Since the best fit is the isochrone that appears most similar to the CMD, many combinations of the parameters can be found as the best fit (minimum $\chi^2$) \citep{dantona18}.

\begin{figure*}
    \centering
    \includegraphics[scale=0.44]{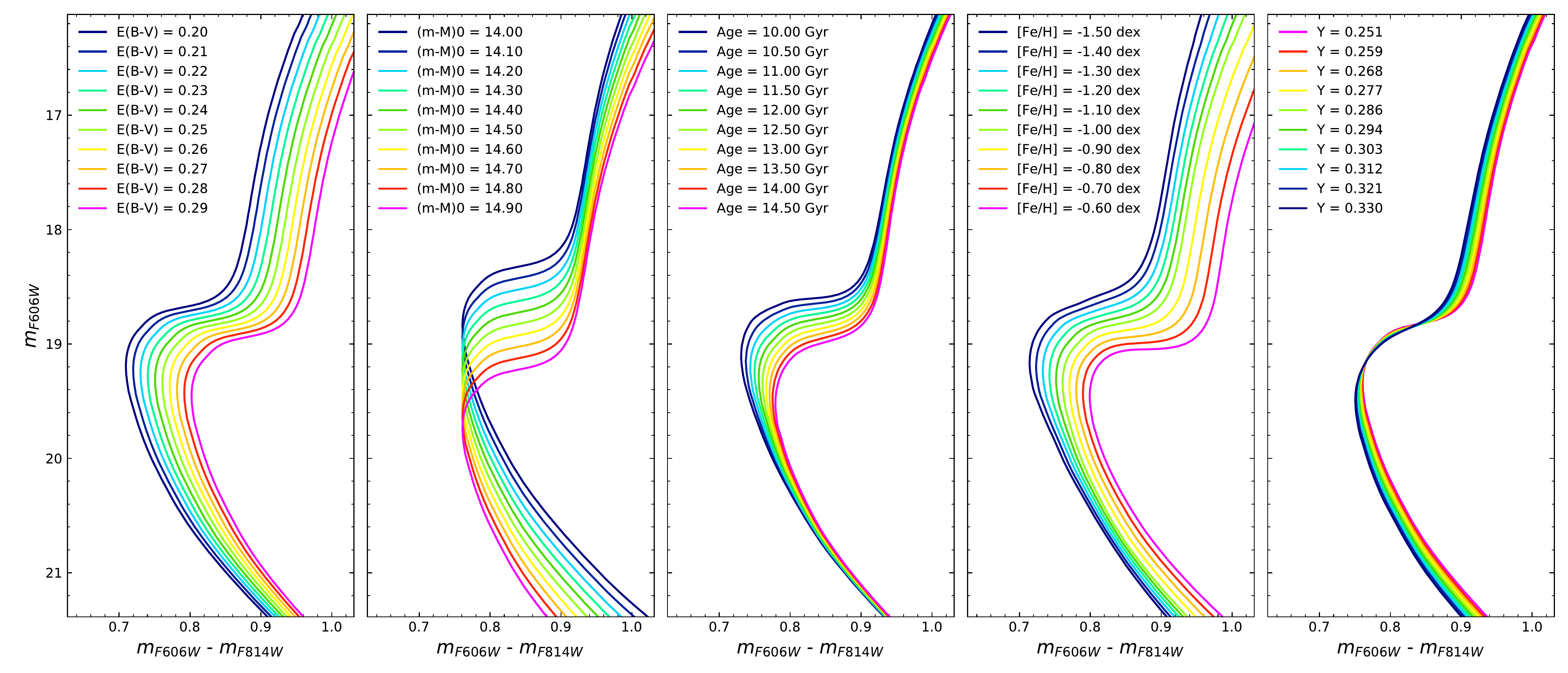}
    \caption{Graphical explanation of how the main five parameters change the morphology and position of the isochrone. The first panel shows the variation due to changes in $E(B-V)$, the second in $(m-M)_0$, the third in Age, the fourth in $\rm [Fe/H]$, and the last one in Y.}
    \label{fig:params}
\end{figure*}

The morphology of the isochrone depends on the age, reddening, 
absolute distance modulus, metallicity, 
and helium abundance. Figure \ref{fig:params} illustrates the  
effects on the shape of isochrones, due to the change in each of these parameters. The reddening $E(B-V)$ changes the location of the isochrone in the diagonal direction because it contributes to the apparent distance modulus $(m-M)_{\lambda}$ and reddening $E(\lambda_1 - \lambda_2)$, without varying the morphology of the isochrone (first panel). For high values of reddening, a second-order correction, from the effective temperature \citep[e.g.][]{ortolani17,kerber19}, has to be taken into account in the isochrone fitting. A vertical displacement is the result of a
change in distance modulus $(m-M)_0$ (second panel). Age $\tau$  affects essentially the position of the turn-off point (TO)
(third panel). The metallicity $\rm [Fe/H]$ has a complex effect on the isochrone, but more strikingly by changing the slope of the RGB,
with a sub-giant branch (SGB) and RGB steeper towards lower metallicities (fourth panel of Figure~\ref{fig:params}). A variation in Y changes the slope of the SGB and the location of the TO, shifting the isochrone to the bluer region of the CMD
(last panel). A review on the interpretation of CMDs
in terms of stellar evolution models can be found in \cite{gallart05}.


\subsection{Bayesian Statistics: Isochrone fitting}

The Bayesian statistics is based on the Bayes Theorem. The probability that two events  (M and D) are 
true, at the same time, according to a null hypothesis H is given by the product probability law:

\begin{center}
  $ \displaystyle P(M,D|H) = P(M|D,H)\times P(D|H)$,
\end{center}    
where $P(M|D,H)$ represents the probability of M to be true if D is true as well according to H, and $P(D|H)$ is the 
probability of D following H. The opposite is also valid:

\begin{center}
   $\displaystyle P(D,M|H) = P(D|M,H)\times P(M|H)$.
\end{center}    

From the hypothesis of the conditional probability of M and D to be the same as D and M, results in the Bayes' theorem:

\begin{center}
   $\displaystyle P(M|D) = \frac{P(D|M)\times P(M)}{P(D)}$,
\end{center}
where, in our case, the evolutionary model is represented by $M$ and the data by $D$.

The posterior distributions $P(M|D)$ are the distributions \emph{a posteriori} of the model (M) and
will give the distributions for each parameter. On the right-hand $P(M)$ are the prior distributions
that give the information \emph{a priori} about the model. The priors are distributions that constrain
the parameters with the physical information.

Assuming that stars are distributed in color and magnitude following a Gaussian distribution and 
disconsidering the dependence of color with magnitude, the likelihood is given by:
\begin{center}
  $\displaystyle P(D|M) = \prod^{N}_{i} \prod^{M}_{j} e^{-\varphi^2_{color}} \cdot e^{-\varphi^2_{Mag}}$,
\end{center}

where $N$ is the total number of the analysed stars and $M$ is the number of points in the isochrone. The $\varphi^2$ is defined as, for example:

\begin{center}
  $\displaystyle \varphi^2_{color_{i,j}} = \frac{1}{2}\left(\frac{color^{obs}_i - color^{iso}_j}{\mathcal{S}_{i} + \sigma^{Cor}_i}\right)^2$,
\end{center}
where $\mathcal{S}$ represents the entropy term of likelihood. This term is responsible for smoothing the region of highest spread and number of stars. The $\mathcal{S}_{i}$, $| {\rm color}_{i}^{\rm obs} - \xi_{\rm f} | $, is calculated for each star by comparison with the fiducial color $\xi_{\rm f}$, which is defined as the median color for a bin of magnitude centered on the magnitude of the $i$-th star.

The maximum likelihood $\mathcal{L}$ corresponds to a maximization of the likelihood function in the parameter space. It is given by (in logarithm form):

\begin{center}
 $ \displaystyle \mathcal{L} = \max\left\{- \sum^{N}_{i=1} \sum^{M}_{j=1} \left[ \varphi^2_{\rm{color}_{\rm i,j}} + \varphi^2_{\rm{Mag}_{\rm i,j}}\right] \right\} $,
\end{center}
 Since the exponential function can reach high values quickly, it is convenient to work with Bayes' theorem in the logarithmic form:
\begin{center}
 $\displaystyle \ln{P(M|D)} = \ln{P(M)} + \mathcal{L} $.
\end{center}

\paragraph{ Priors}
The prior distributions ($P(M)$) are the main difference between
the Bayesian and the frequentist statistics. These distributions impose constraints on the free parameters, restricting the set of parameters to be explored. In an isochrone fitting, these priors reflect the physical constraints, such as: \textit {(a)} the upper age limit as the age of the Universe \citep{planck16}; \textit{(b)} the metallicity values taken from high-resolution spectroscopy; \textit{(c)} distances constrained and primordial He content from RR Lyrae mean magnitudes, for example; and \textit{(d)} non-negative reddening values.

\paragraph{ Marginalization}

In order to explore the parameter space as a whole and to get the posterior distributions of each parameter, we applied the Bayes' theorem with the Metropolis-Hastings (MH) algorithm \citep{metropolis53, hastings70}. The method is basically an exclusion iterative algorithm, built firstly to solve problems of statistical physics. The MH method compares the random probabilities trying to reach the minimum energy state, which justifies that we can neglect the normalization term of the Bayes' law. The final result of MH is a chain with $n$ energies for $m$ states that is known as Markov chain. For the applications with random distributions, which means Monte Carlo methods, the result from the MH algorithm is called Markov chain Monte Carlo \citep[MCMC,][]{hogg18}. To get the probability distributions of the parameters, the marginalization is executed by the integral:

\begin{center}
    $\displaystyle \mathcal{P}(\overrightarrow{\phi}) = \int \mathcal{L}(\overrightarrow{\phi})\times p(\overrightarrow{\phi})\, {\rm d}\overrightarrow{\phi} $,
\end{center}
where $(\overrightarrow{\phi})$ represents the parameter space. To perform the marginalization from MH algorithm and MCMC method, we employed the \texttt{Python} library \texttt{emcee} \citep{foreman-mackey13}. 

\subsection{Multiple Stellar Populations in GCs}\label{subsec:mptagging}

\begin{figure*}
    \centering
    \includegraphics[scale=0.53]{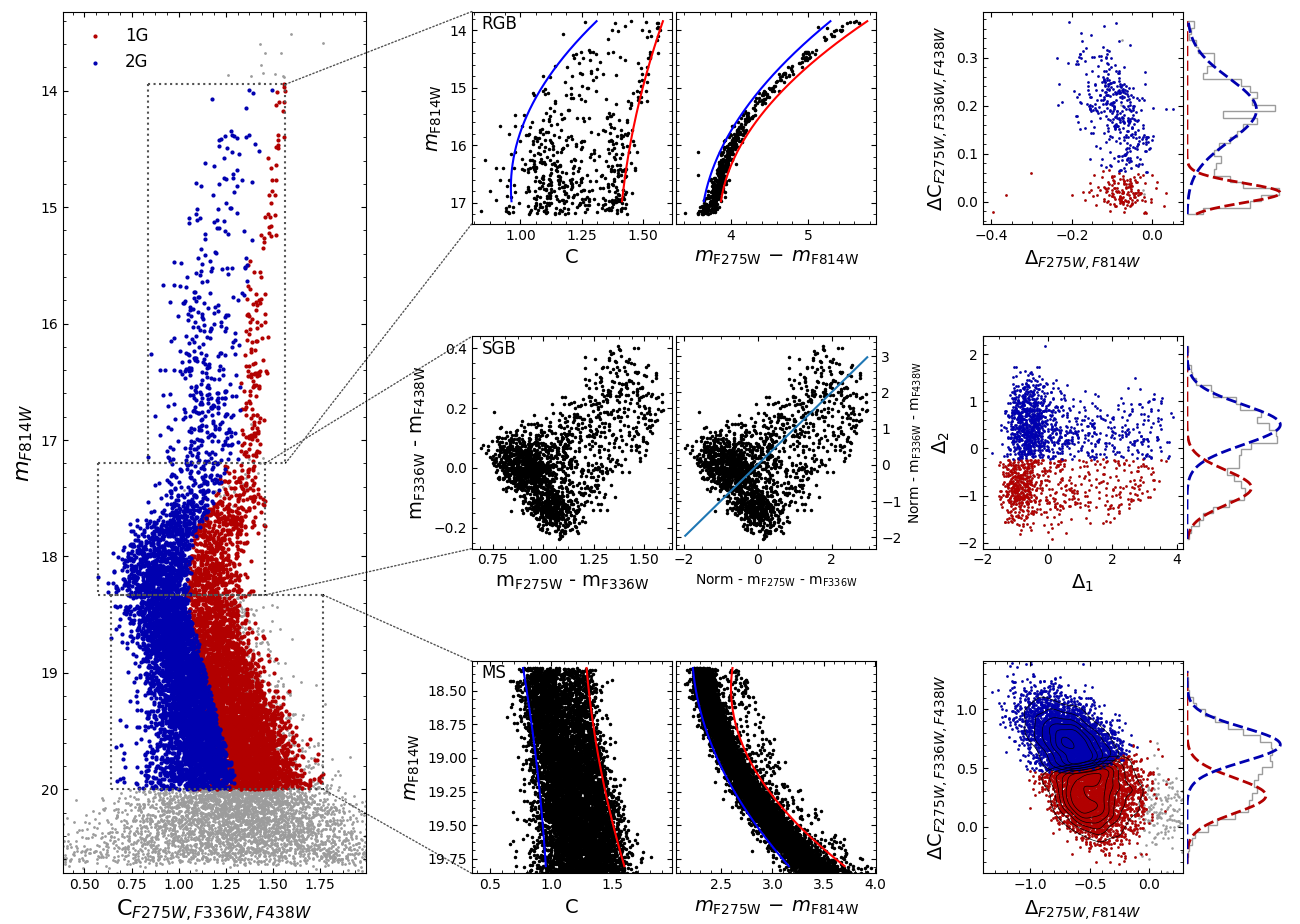}
    \caption{ MP separation and population tagging applyed to synthetic data with $\Delta \tau = 0.5$ Gyr. Left panel shows the pseudo-color C, which gives a pronounced MP separation. Middle panels show the procedure we apply to separate the stellar populations, from top to bottom are the RGB, SGB, and MS stars, respectively. Right panels show the stars identified to belong to the 1G and 2G.}
    \label{fig:msp-do}
\end{figure*}

Before carrying on the analysis of MPs, in this section we describe the separation of stellar populations in the CMDs. The stellar population tagging allows us to distinguish the first (1G) and second (2G) generation stars (and subsequent ones) from a given CMD. Figure \ref{fig:msp-do} shows the procedure we follow to separate the stellar populations in each region of the created synthetic CMD with $\Delta \tau_{\rm 1G,2G} = 0.50$ Gyr. We adopted a Dartmouth (DSED) isochrone with [Fe/H]$=-1.26$, $E(B-V)=0.18$, $(m-M)_0=14.38$, and $\tau=13.0$ Gyr.


In Milone et al. (2013) the pseudo-color C was defined, with the purpose to maximize the separation among MPs on the CMD. Piotto et al. (2015) have shown the power of {\it HST} UV filters F275W, F336W, and F438W  to separate the MPs. F275W is sensitive to OH and F438W to CN and CH. For these filters, the 1G stars are fainter than the 2G because the latter are oxygen- and carbon-poorer than the 2G ones. For the filter F336W, which is sensitive to NH, the 1G stars are brighter than the 2G stars, given the fact that the 2G stars are nitrogen-richer. 
Note that stronger lines lead to larger opacity, and lower brightness. For these reasons, the color (F275W-F438W) inverts the stellar populations on the CMD with respect to the color (F336W-F438W). In that color, the 2G stars seem to be redder than the 1G stars (Piotto et al. 2015, their Figure 2).


\paragraph{Chromosome maps (RGB and MS)}
\citet{milone17} describe the method of MP separation using 
chromosome maps based on combinations of UV HST filters.
\citet{lee19} used  UBV data to distinguish MPs, and
reviewed methods discussed earlier. To construct the chromosome map diagrams, we adopt the method presented in \citet{milone17} that is briefly described below. For the CMDs m$_{\rm F814W}$ vs. C$_{F275W,F336W,F438W}$ and m$_{\rm F814W}$ vs. (m$_{\rm F275W} - \rm m_{\rm F814W}$), the red and blue fiducial lines are defined by $96^{th}$ and $4^{th}$ percentiles, respectively. The top- and bottom-middle panels of Figure~\ref{fig:msp-do} show the red and blue fiducial lines enclosing the RGB and MS stars, respectively. The axis of chromosome map are the relative distance between each stars and the fiducial lines, defined by: 


\begin{center}
 $ \displaystyle \Delta_{C\,F275W,F336W,F438W} =  \frac{C_{\rm r} - C}{C_{\rm r} - C_{\rm b}} $,
\end{center}

\begin{center}
 $ \displaystyle \Delta_{F275W,F814W} =  \frac{G - G_{\rm r}}{G_{\rm r} - G_{\rm b}} $,
\end{center}
where the indices $r$ and $b$ refer to the red and blue fiducial lines, respectively. The color $G$ represents m$_{\rm F275W} - \rm m_{\rm F814W}$.

The diagram $\Delta_{C\,F275W,F336W,F438W}$ vs. $\Delta_{F275W,F814W}$ quantifies the color distance of each star to the blue and red envelopes, so that the $\Delta$-value is 
closer to zero as the star is closer to the red envelope. The right panels of Figure~\ref{fig:msp-do} show the final chromosome maps for the RGB (top) and MS (bottom), respectively, for the synthetic CMD. 

Some modifications on the identification of the MPs were implemented in the
original method from \cite{milone17}, in order to preserve a uniformity in the MPs separation for the three evolutionary stages (MS, SGB, RGB). The identification of the MPs is done using the Gaussian Mixture Models (GMM), that is a non-supervised machine learning algorithm, which searches to fit $K$ Gaussian distributions to a sample of $N$ data. The fit comes from the basic equation of the Bayes' theorem:

\begin{center}
 $ \displaystyle G(x) = \sum_{i=1}^{K} \phi_i \times \mathcal{N}\left( x\, |\, \mu_i, \sigma_i \right) $,
\end{center}
where $\mathcal{N}( x\, |\, \mu_i, \sigma_i )$ represents the ith Gaussian distribution with mean of $\mu_i$  and standard deviation of $\sigma_i$. This algorithm was adopted from the \texttt{python} library \texttt{Scikit-learn} \citep{pedregosa11}.

We here assume two subclasses for GMM in a two-dimensional plane. Then, each star is classified as 1G or 2G according to the strength of the two Gaussian distributions on that point of the chromosome map. The separation between the two {populations} includes clear members of both, but as well stars in the limiting intersection, that can contaminate each other samples. This analysis can be improved by increasing the number of subdivisions in GMM to select the bona-fide stars of each stellar populations, as in \cite{milone18}.

\paragraph{Two-color diagrams (SGB)}
Since the SGB sequence, depending on the adopted filter and the metallicity of the cluster, could be nearly horizontal and their MPs could appear mixed, the chromosome maps are not effective with these stars. Therefore,
we applied a conventional two-color diagram $m_{\rm F336W}-m_{\rm F438W}$ vs.
$m_{\rm F275W}-m_{\rm F336W}$, as described in \citet{nardiello15b}. In order
to apply the GMM procedure (same as described in the previous section), $\Delta_1$ and $\Delta_2$ are the axes that were
normalized and then rotated counterclockwise by an angle of 45$^{\circ}$. The method is graphically represented in Figure~\ref{fig:msp-do} (middle panels).

\subsection{Age difference $\Delta\tau$ }\label{sec:deltatau}

The origin of the 2G (and subsequent {populations}) stars is a major challenge in the MP analyses. Most scenarios  trying to explain  MP formation predict an age difference  ($\Delta \tau$) between the first and the later {populations} \citep{bastian18}. For example, the scenario of Asymptotic Giant Branch (AGB) stars polluting the second and subsequent {populations},
predicts a difference around 100 Myr \citep[][]{dantona16}, up to 200-700 Myr from the delay of X-ray binaries \citep{renzini13, renzini15}. Another scenario is that of the supermassive stars (SMS). Multiple stellar {populations} can be formed from multiple bursts of SMSs with intervals of a few Myr \citep{gieles18}. Another possibility are the fast rotating massive stars (FRMSs) that would enrich the interstellar medium in about 40 Myr \citep[][]{decressin07,krause13}. Therefore, the age difference between the first and next {populations} is an important parameter to give hints to their plausible origin.

From our population tagging method, we can analyse separately each {stellar population} from their CMDs. To perform the isochrone fitting in the context of MPs we developed a hierarchical algorithm to estimate the $\Delta \tau$ between the first and subsequent {populations}. The hierarchical algorithm considers the stars as a SSP
first, and subsequently each {stellar population}. For a SSP we leave all parameters free. In the context of MPs, it is expected that the age of a SSP is a weighted average age  of each stellar population. Consequently, for the example of two stellar {populations}, the ages could be derived from:

\begin{center}
$ \displaystyle
\begin{array}{l c l}
    \tau_{\rm 1G} & = & \tau_{\rm SSP} + \Delta \tau \times \left({\rm N}_{\rm 1G}/{\rm N}_{\rm total}\right),  \\
    & & \\
    \tau_{\rm 2G} & = & \tau_{\rm SSP} -  \Delta \tau \times \left(1 - {\rm N}_{\rm 1G}/{\rm N}_{\rm total}\right).  \\
     & 
\end{array}
$
\end{center}

The hierarchical method fits the first {population} and applies the constraints of distance, reddening, and metallicity to the second (or subsequent) one(s). Hence, the procedure to compute the $\Delta \tau$ turns out simply to be $\Delta \tau = \tau_{\rm 1G}-\tau_{\rm 2G}$. This procedure considers that 1G stars were formed earlier than others, which is logical when our objective is to estimate a $\Delta \tau$. The likelihood of hierarchical procedure $\ln{\rm P(M|D)}$ takes into account the constraints of a stellar cluster as a whole. For example, all stars must have  the same values of distance and must be influenced by interstellar dust in the same way. Therefore, the likelihood of 1G ($\mathcal{L}({\rm 1G})$) and NG ($\mathcal{L}({\rm NG})$) are dependent on the likelihood of SSP ($\mathcal{L}_{\rm SSP}$). The total likelihood $\ln{\rm P(M|D)}$ is a linear combination of the priors and the likelihood 
of  each {stellar population} with influence of SSP parameters:

\begin{center}
 $\displaystyle \ln{\rm P(M|D)} = \ln{\rm P(M)} + \sum_{i=1}^{N} \left[ \mathcal{L}({\rm \, [i]G\, })_{\rm SSP} + \ln(f_{[i]G}) \right] $.
\end{center}
where $f_{[i]G}$ represents the fraction of stars that belong to the $i$-th population. A similar likelihood based on MPs and weighted by the fraction of stars is applied in \citet{ramirez-siordia19}.

Here, we are adopting that the 1G stars have primordial helium content (Y), which is consistent with the literature \citep{bastian18}. \citet{wagnerkaiser16}  performed a bayesian isochrone fitting, in the context of MPs, for a sample of 30 GCs. Differently from the present work, they fitted the value of Y for the 1G stars, resulting in some cases in a high content of $Y_{1G} \sim 0.30$. They also assumed the same age for both analysed stellar populations. On the contrary, we are interested in finding if there is an age difference between the stellar populations. Even though our approach is similar to the one applied in \citet{wagnerkaiser16}, the methods are based on different assumptions.

\section{Controlled Experiment}\label{sec:tests}

\begin{figure*}
    \centering
    \includegraphics[scale=0.59]{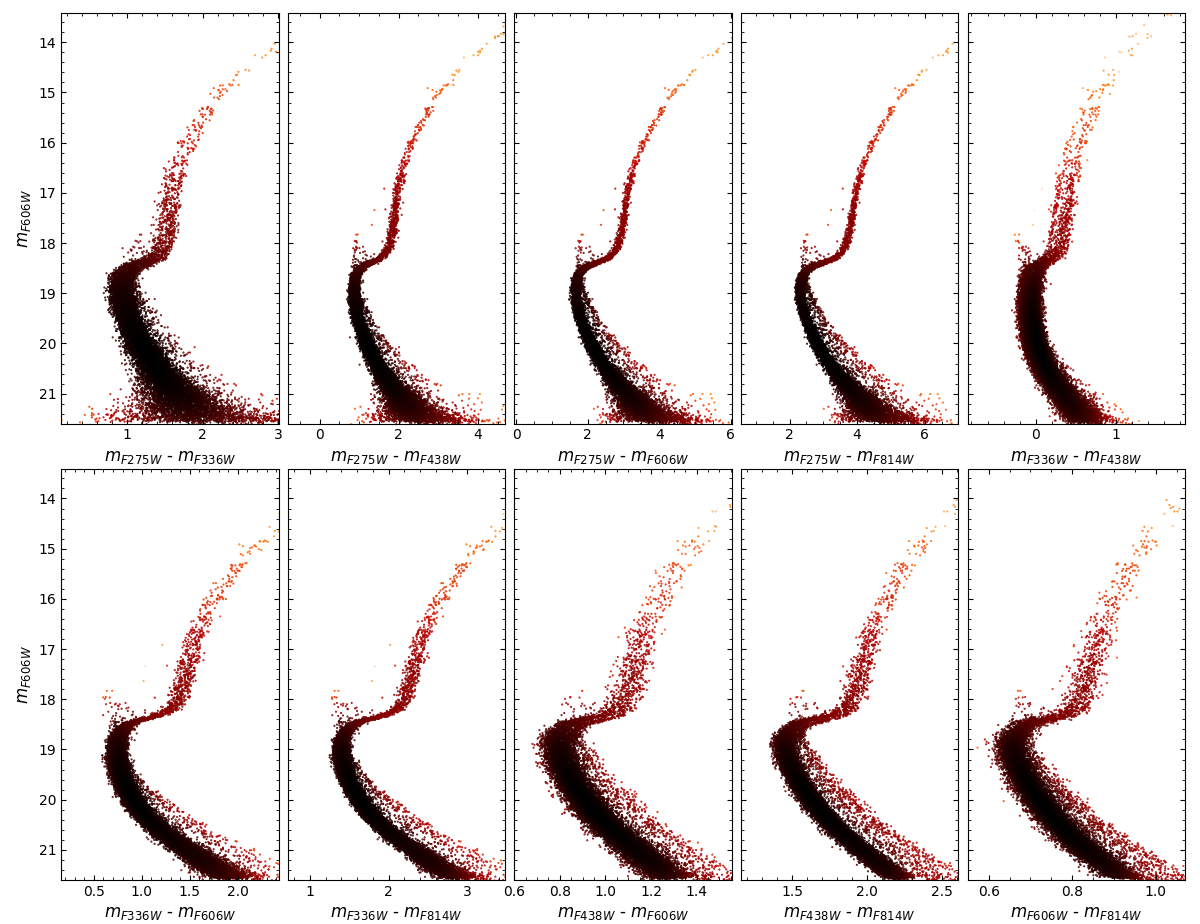}
    \caption{CMDs for the Synthetic Data using  a DSED isochrone with age $= 13.0$ Gyr, [Fe/H]$ = -1.26$, $E(B-V) = 0.18$, $(m-M)_0 = 14.38$, $\Delta \tau = 0.50$ Gyr and fraction of 1G stars ($N_{1G}/N_{total}$)$= 0.360$, generated from {\it HST} filters. All available combinations of filters are shown.}
    \label{fig:cmd-mosaic}
\end{figure*}

In this Section, we test the reliability of our analysis by using synthetic CMDs.
First, we constructed a synthetic CMD using an error function obtained from the atlas extracted by \citet{nardiello18} from the data of the {\it HST} UV-Legacy Survey of Galactic Globular Clusters \citep{piotto15}, allowing us to simulate MPs with the synthetic data. The stellar evolutionary model adopted was the DSED isochrone with Z $\sim$ 0.002 with [$\alpha$/Fe] = +0.4, and age of 13.0 Gyr, as reported in Table~\ref{tab:my_label}, corresponding to typical values of moderately metal-poor bulge GCs \citep[e.g.][]{kerber18,kerber19}. We simulated the CMD of a cluster with a total number of $10,000$ stars (N$_{total}$) that host ~36\% of 1G stars with an age of 13.0 Gyr and 64\% of 2G stars 0.5 Gyr younger than 1G stars. We considered a fraction of binaries (f$_{\rm bin}$) of $30 \%$ and a minimum mass ratio (q$_{\rm min}$) of $0.60$. Resulting CMDs combining the different available filters are shown in Figure~\ref{fig:cmd-mosaic}.

\input{TAB-MockParams.tex}

\subsection{Sources of uncertainty}

In our method, during the isochrone fitting, we compute the likelihood star-by-star. To keep the high performance of MCMC, we imposed a range in magnitudes based on stellar evolutionary models. The third panel of Figure~\ref{fig:params} shows that there is no significant difference regarding the age for the $\sim 3$ magnitudes brighter than the TO. For this reason, we do not take into account stars above this limit in the likelihood calculation. 

The faintest stars are limited to the completeness limit, meaning that the number of faint stars depends on the photometric depth. There are no differences between the isochrones in the databases employed in \texttt{SIRIUS} for the faintest stars ($\sim 2$ magnitudes below the TO), therefore the fit does not depend on the faintest stars. \cite{ramirez-siordia19} presented an analysis considering the faintest stars. They concluded that the effect of faintest stars only increases the uncertainties without changing the mode of distribution, since the isochrones do not seem to be different for the faintest stars, as shown in Figure~\ref{fig:params} (third panel).

As regards binary stars, their magnitudes represent the combination of the fluxes from the two companion stars. Since the magnitude is the logarithm of the stellar flux, for a binary system with two stars of the same mass, the magnitude of this system corresponds to the magnitude of one star subtracted by $2.5\times \log(2) \sim 0.75$ \citep{kerber02, kerber07}. The decrement in magnitude tends to have the binary stars to be brighter and redder on the CMD. To reduce the effect of binary systems during the isochrone fitting, \texttt{SIRIUS} takes into account only the stars within $3\sigma$ from the fiducial line of the CMD.

The standard BaSTI isochrones  overestimate ages by $\sim 0.80$ Gyr, with
respect to DSED isochrones. The main reason for this discrepancy is that BaSTI isochrones do not include atomic diffusion in the calculations, 
among other differences in basic physics. Whereas the solar alpha-to-iron more complete models, including atomic diffusion are already
available in \citet{hidalgo18}, the available alpha-enhanced models taking this effect into account are not yet available.



\subsection{Sanity Check}

In the optical wavelengths some filters are  more sensitive to some properties than others. For the NIR  filters the effect of interstellar medium extinction is considerably lower than for the UV filters. Also, a color combining filters with a small band width is more suitable to observe the structures on the CMD. Therefore, the combination of magnitudes and colors on the CMD is very important regarding the information that is expected to be obtained from isochrone fitting. In order to estimate the effect of the choice of color we performed the isochrone fitting using ten different colors, without spreading the stars, combining the five {\it HST} filters available in the
UV Legacy survey of globular clusters \citep{piotto15}.

\begin{figure*}
    \centering
    \includegraphics[scale=0.45]{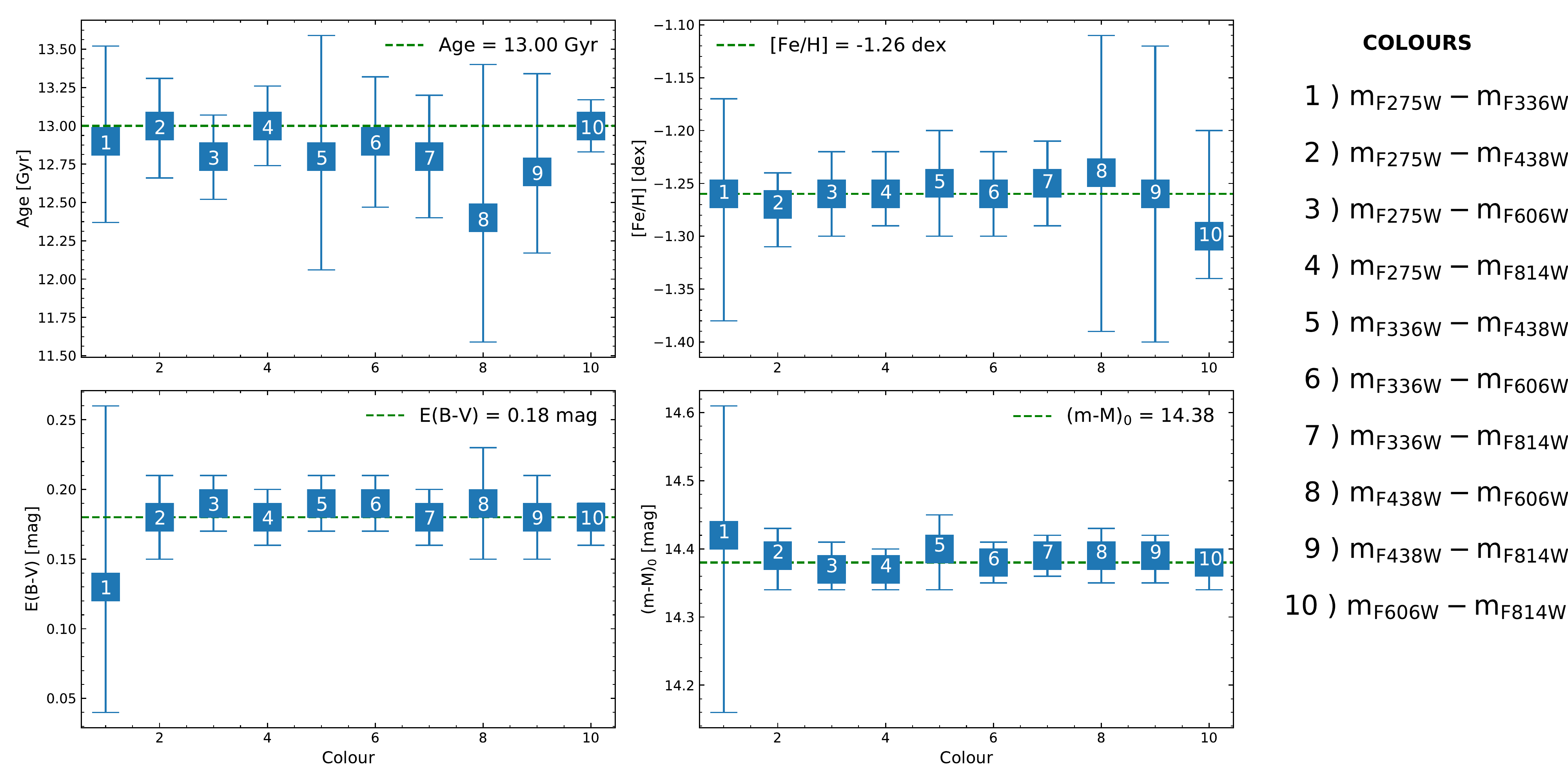}
    \caption{Sanity check with no-spread data, the parameter space as function of color. The posterior distributions of each parameter for the ten combinations of {\it HST} filters of the UV Legacy survey of globular clusters \citep{piotto15}. DSED isochrones are adopted. The numbers represent each color.}
    \label{fig:sanity-results}
\end{figure*}

Firstly, we perform the fit considering the SSP without taking into account the photometric spread of stars. The DSED isochrones are here fitted to the synthetic No-Spread catalogue data (Table~\ref{tab:my_label}) with the purpose of checking if the input parameters of the synthetic CMD are recovered. For this test, we adopted uniform distribution priors for all parameters. The range of values we used are: for age, between 10 to 15 Gyr; for the metallicity, between 0.00 to $-2.00$ dex; for reddening, between 0.0 to 1.0 mag; and for the distance modulus, between 12.0 to 16.0 mag. Figure~\ref{fig:sanity-results} shows the behavior of the parameter space as a function of color. It can be observed that the age is the most sensitive parameter to the filters, whereas the other parameters vary only slightly with the choice of filters. For color 8 (third lower panel in Fig. \ref{fig:cmd-mosaic}), which is equivalent to B-V, there is a strong effect on the age, whereas for color 6
(first lower panel in Fig. \ref{fig:cmd-mosaic}) the parameters are closer to the original ones. Color 10 (m$_{\rm F606W}-{\rm m}_{F814W}$, last lower panel in Fig. \ref{fig:cmd-mosaic}), is also close to the input values and has small uncertainties due to its lowest reddening-dependency. Therefore, for our analysis, we chose  color 10.


\begin{figure}
    \centering
    \includegraphics[scale=0.5]{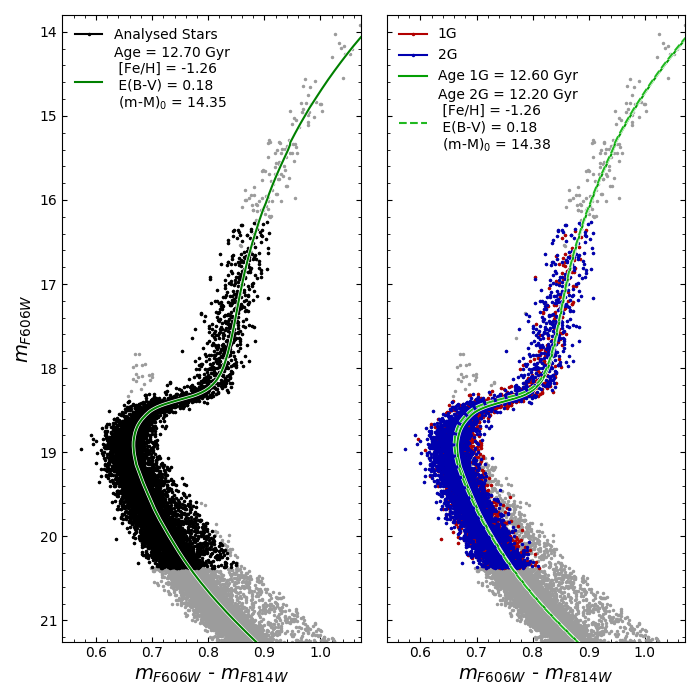}
    \caption{Sanity check with spread data, isochrone fitting for the synthetic CMD considering SSP (left) and MPs (right) for DSED isochrones. The grey dots are discarded for the fit.}
    \label{fig:syn_fits}
\end{figure}

\begin{figure*}
    \centering
    \includegraphics[scale=0.5]{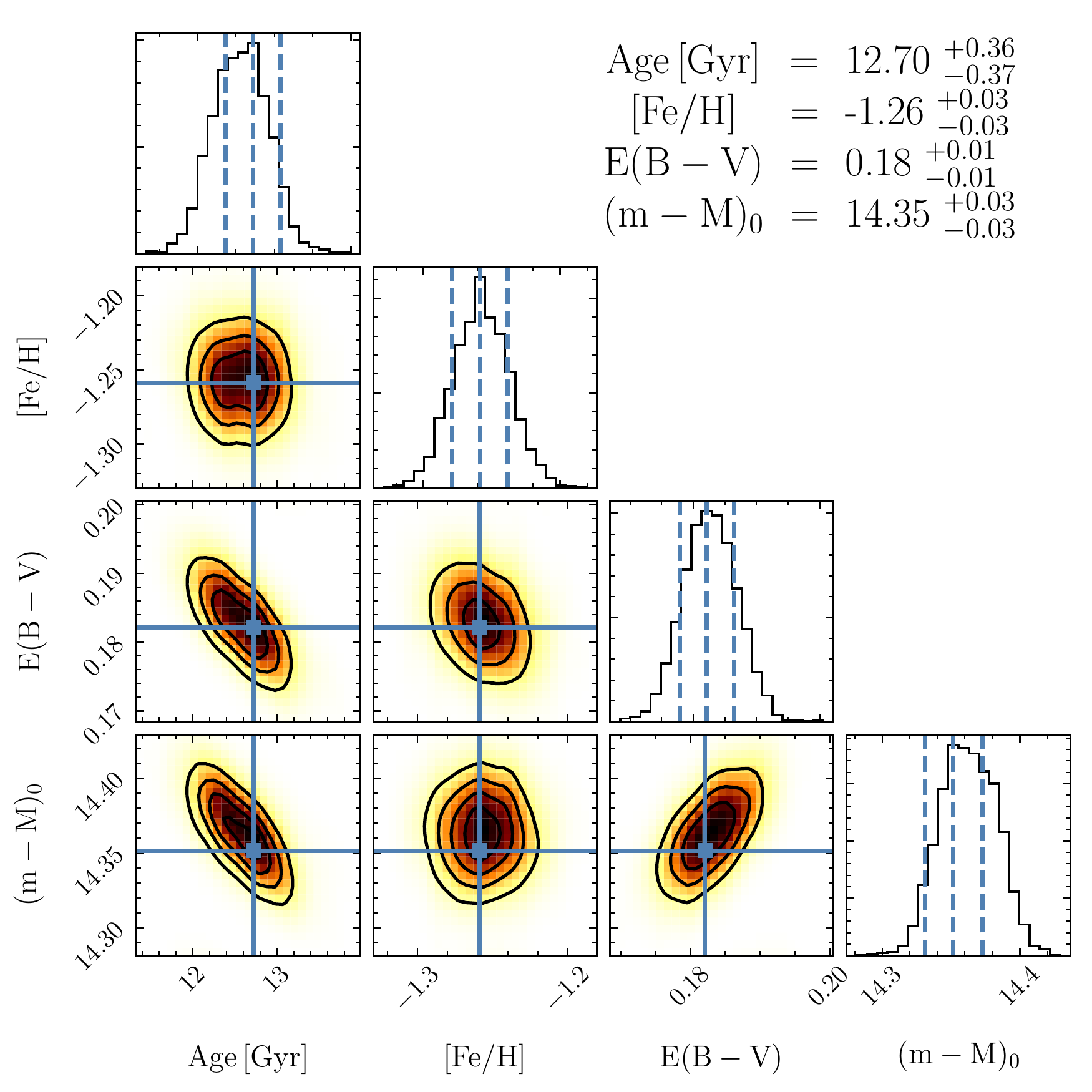}
    \includegraphics[scale=0.5]{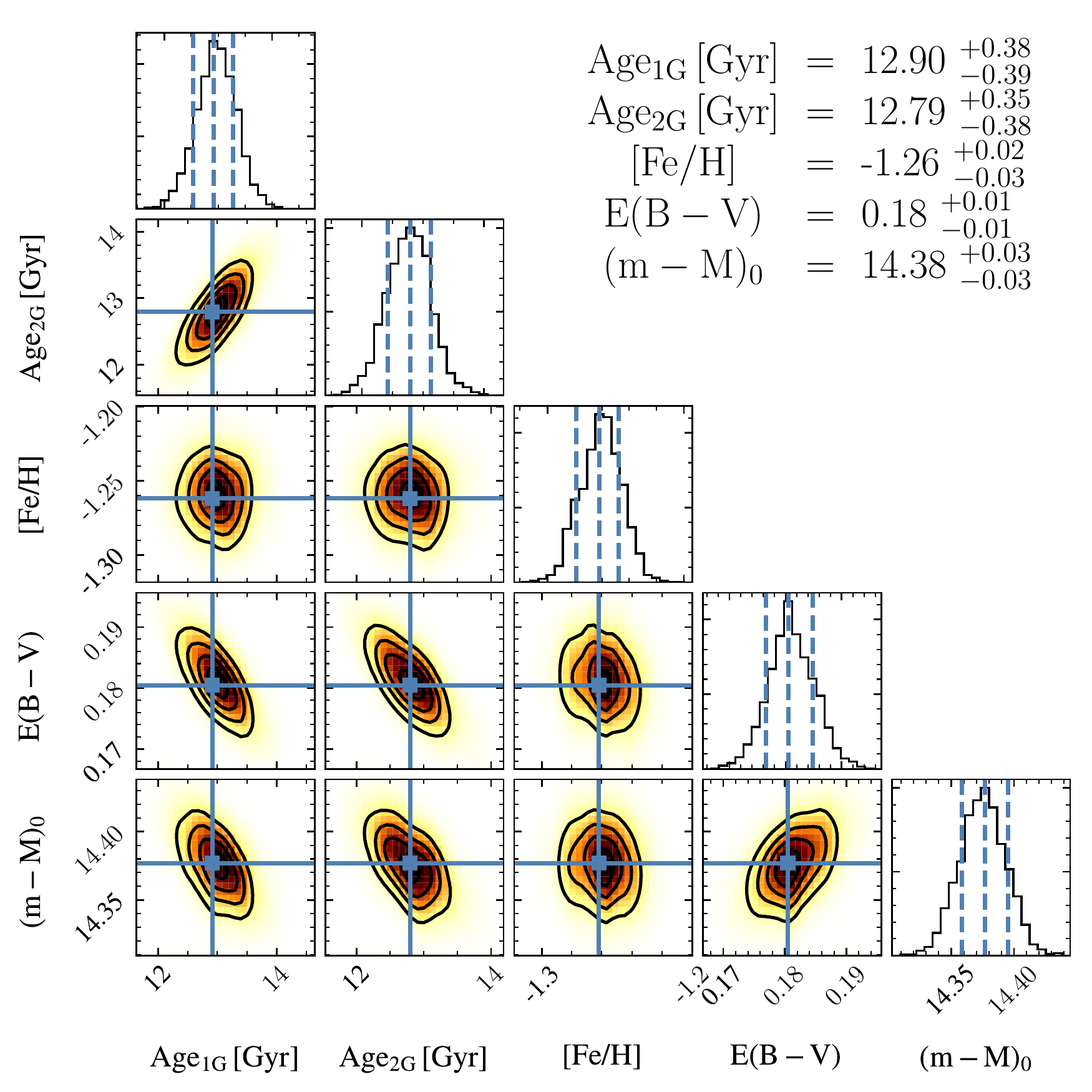}
    
    \includegraphics[scale=0.5]{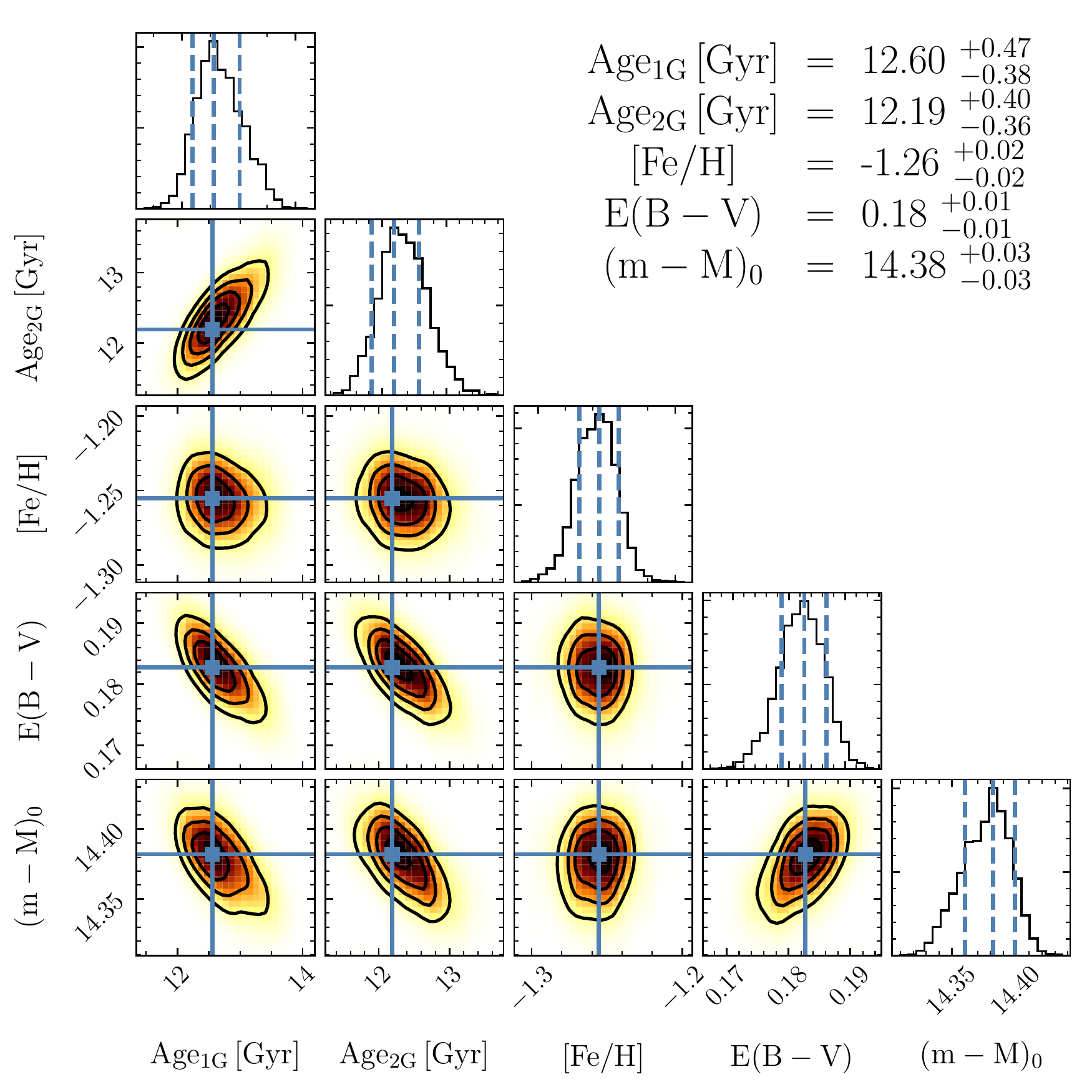}
    \includegraphics[scale=0.5]{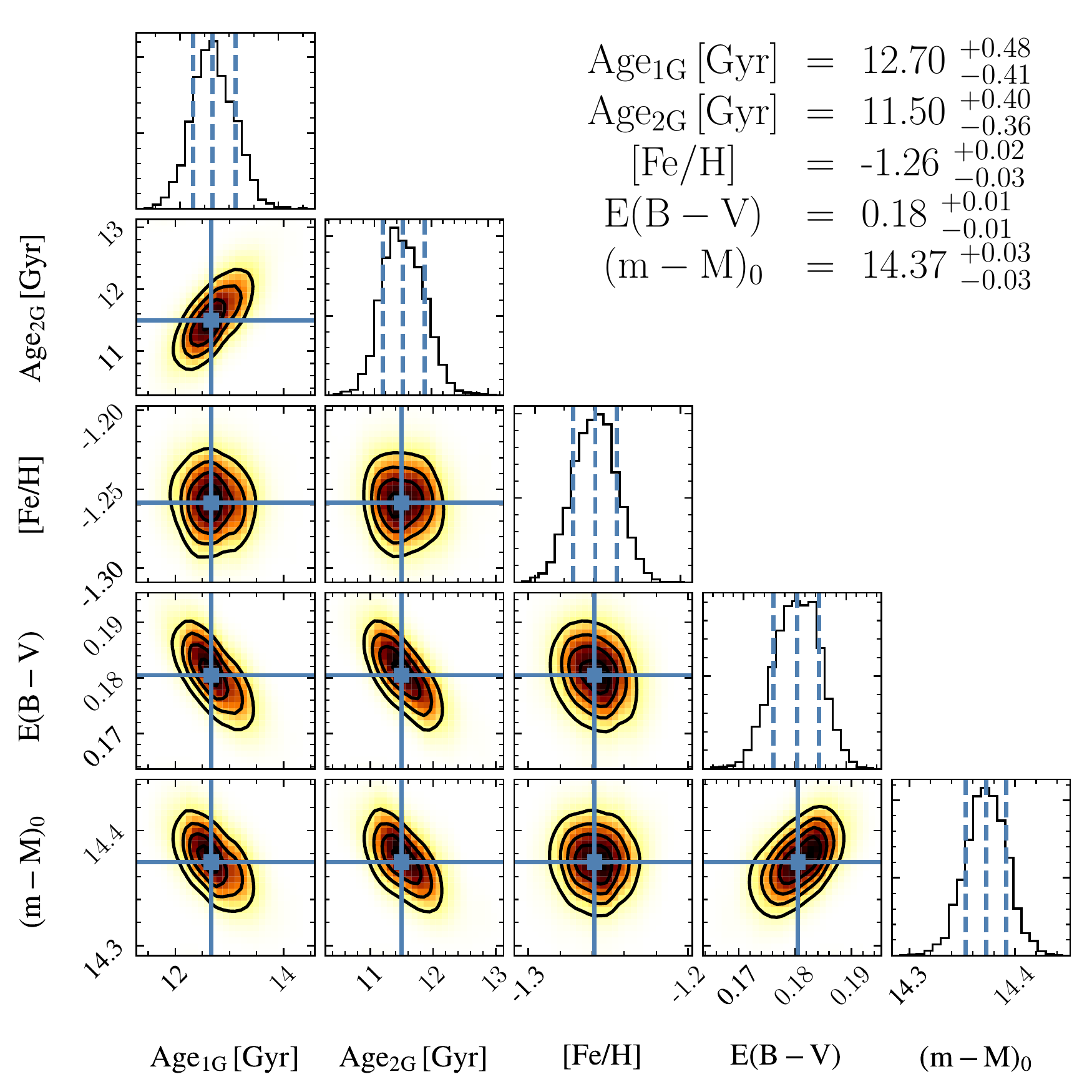}
    
    \caption{Sanity check 2, corner plots using DSED isochrones, relating physical parameters. Top left panel: results of the sanity check applied to a synthetic SSP CMD
    where Monte Carlo spread of data is implemented, with a $\Delta \tau = 0.50$ Gyr. Other panels: 1G and 2G combined for $\Delta\tau = 0.10$ Gyr (top right), $\Delta\tau = 0.50$ Gyr (bottom left), and $\Delta\tau = 1.50$ Gyr (bottom right). 
    }
    \label{fig:sanity-dsed}
\end{figure*}

Secondly, to verify the sensitivity of the method, we simulate real data through synthetic CMDs
to perform the isochrone fitting, taking into account a spread of stars, and assuming Gaussian priors centered on the parameters given in Table \ref{tab:my_label} (Spread). In Figure \ref{fig:syn_fits}, we show the isochrone fitting for the synthetic CMD with $\Delta \tau = 0.50$ Gyr, assuming that it is SSP (left panel) and MPs (right panel). 
We employ the corner-plots to present the posterior distributions 
(Figure \ref{fig:sanity-dsed}). They show the $N$ parameter space in a 2D representation, where it is possible to see the correlations between the parameters. As the best value for each parameter we adopted the mode of the distributions. For the confidence interval, we selected the 16$^{\rm th}$ and 84$^{\rm th}$ percentile of the distributions that give us the values inside $1\sigma$ from the mode. The top-left panel in Figure \ref{fig:sanity-dsed} shows the corner-plot for the DSED SSP isochrone fitting. 
Figure~\ref{fig:sanity-dsed}, in the top-right, bottom-left, and bottom right panels show the results for the age derivation in the context of MPs using DSED. 

\input{TAB-ResSyn.tex}

Even though the spread of stars changes the visual aspect of the CMD, the parameters obtained from the isochrone fitting given in Table \ref{tab:sanity2} for SSP and MPs, are both in good agreement with the input values from Table~\ref{tab:my_label}.
In conclusion, in this section we were able to describe the approach and check the validity of \texttt{SIRIUS} in the context of MPs. 

\section{Application to the halo globular cluster NGC~6752}\label{sec:app}

{\it HST} photometric data for NGC~6752 in the ultraviolet (UV) filters within the UV-Legacy Survey GO-13297 (PI. G. Piotto), and in the optical within GO-10775 (PI. A. Sarajedini) are used. These programs made available data in the UV filters F275W, F336W, and F438W from the Wide Field Camera 3 (WFC3), and the optical filters F606W and F814W from the Wide Field Camera of the Advanced Camera for Survey (WFC/ACS). The newly reduced catalogs presented in \citet{nardiello18}
are used.

NGC 6752 is a halo cluster, located at l = 336$^\circ$49, b = -25$^\circ$63,
with a distance from the Sun d$_{\rm \odot}$ = 4.0 kpc \citep[][edition 2010]{harris96}\footnote{www.physics.mcmaster.ca/~harris/mwgc.dat}.
A metallicity of [Fe/H]$=-1.48\pm0.07$ dex was derived by \citet{gratton05} from high resolution spectroscopy ($R = 40,000$) of seven stars near the red giant branch bump. \citet{gratton03} and \citet{vandenberg13} obtained an age of $12.50\pm0.25$ Gyr and $13.4\pm1.1$ Gyr, respectivaly. 
\citet{carretta12} identified three stellar populations based on three values of abundances of O, Na, Mg, Al, and Si elements that are sensitive to stellar {populations} in GCs, denominated as first (P), intermediate (I), and extreme (E) populations. \citet{milone13} gave the first photometric evidence of three stellar populations by using \textit{HST} data. \citet{nardiello15a}, using FORS2/VLT data, have observed the split of the MS of NGC~6752 using UBI filters, and calculated the radial distribution of the populations and the difference in helium between the 1G and 2G stars. \citet{milone19} confirmed  the existence of three stellar populations from NIR photometric data on MS stars. \citet{cordoni19} analysed the kinematics of the P and E populations of NGC~6752, and they found that there is no difference in rotation between the two stellar populations.  

\begin{figure*}
    \centering
    \includegraphics[scale=0.42]{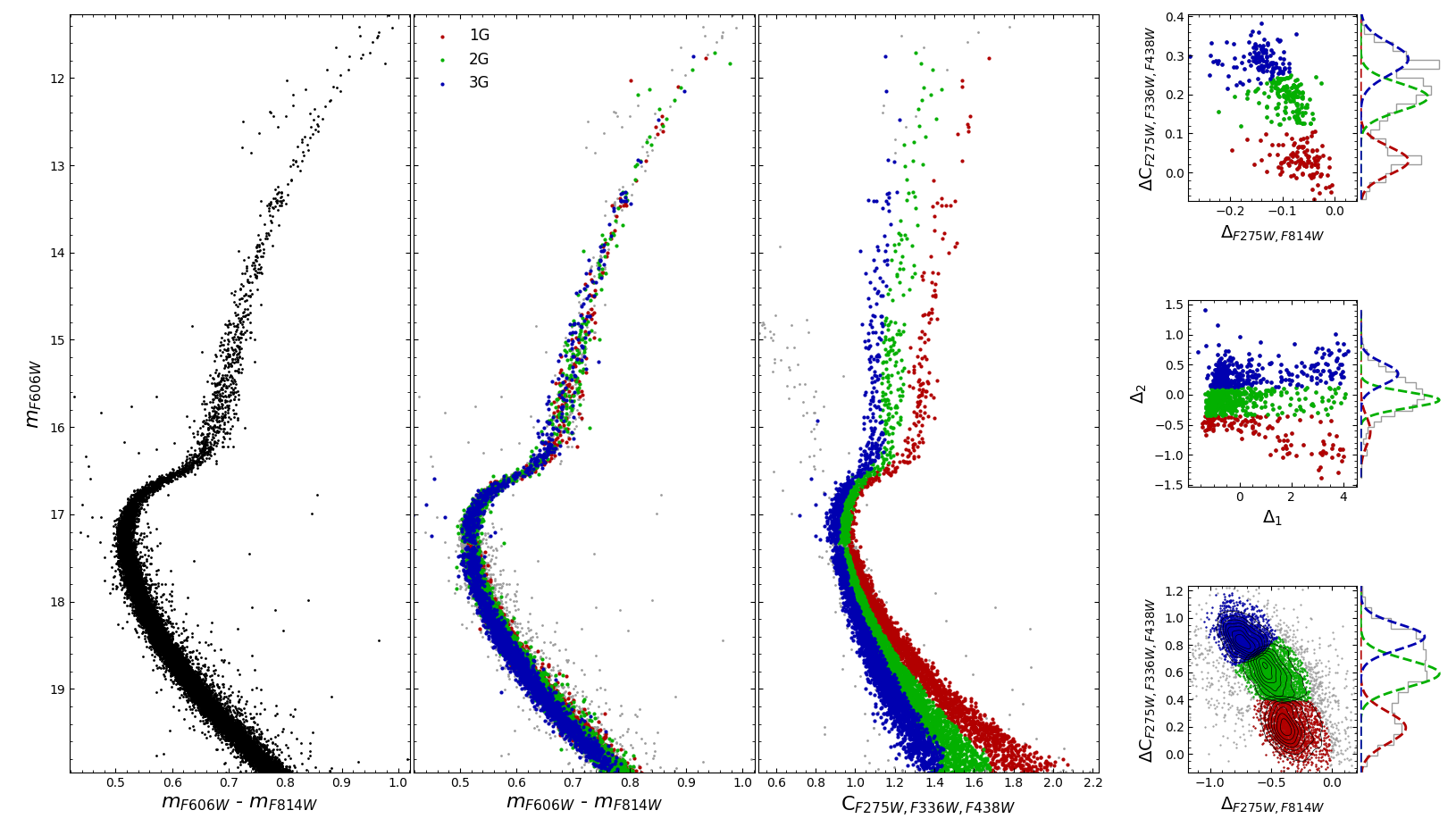}
    \caption{Multiple stellar populations in NGC~6752. Left panel:
    SSP; Middle panel: same as left panel, but color-identified stars;
    Right panel: pseudo-color showing the clear separation of three
    stellar {populations}.}
    \label{fig:ngc6752_mps}
\end{figure*}

In order to separate the populations P, I, and E (hereafter 1G, 2G, and 3G), the number of components on GMM were increased to three for the RGB and SGB, and to four for the MS. The classification of 1G, 2G, and 3G stars is in agreement with \citet{milone13}, since a clear distinction of three stellar populations
can be verified in Figure~\ref{fig:ngc6752_mps}. \citet{milone13} derived the mass fraction of each population to be of $\sim 25$, $\sim45$, and $\sim30$ per cent, respectively. We found a fraction of stars of $25\pm5$, $46\pm7$, and $29\pm5$ per cent for the 1G, 2G, and 3G, respectively, in excellent agreement with \citet{milone13}. 

\begin{figure*}
    \centering
    \includegraphics[scale=0.57]{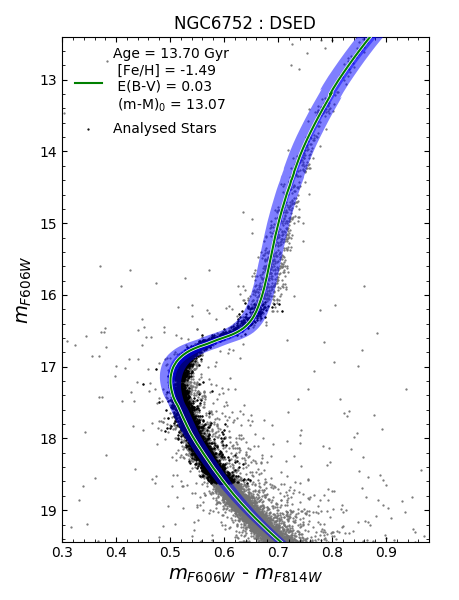}
    \includegraphics[scale=0.5]{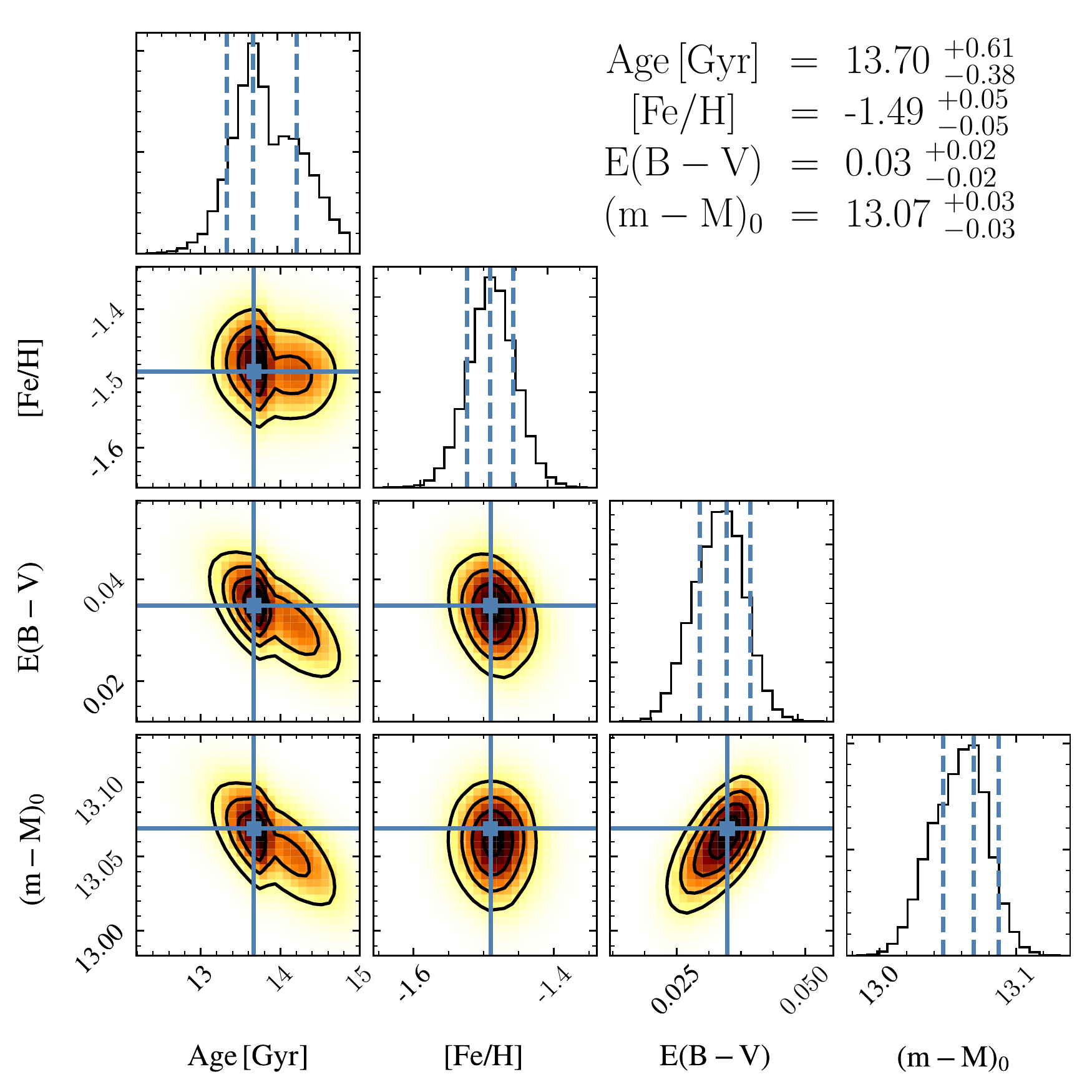}
    \caption{ Results for the SSP analysis of NGC~6752. Left panel: CMD with the result from isochrone fitting, green line is the most probable solution, and the blue strip is the solutions within 1$\sigma$. Right panel: The posterior distributions.
    }
    \label{fig:ngc6752-dsed_ssp}
\end{figure*}

\begin{figure*}
    \includegraphics[scale=0.5]{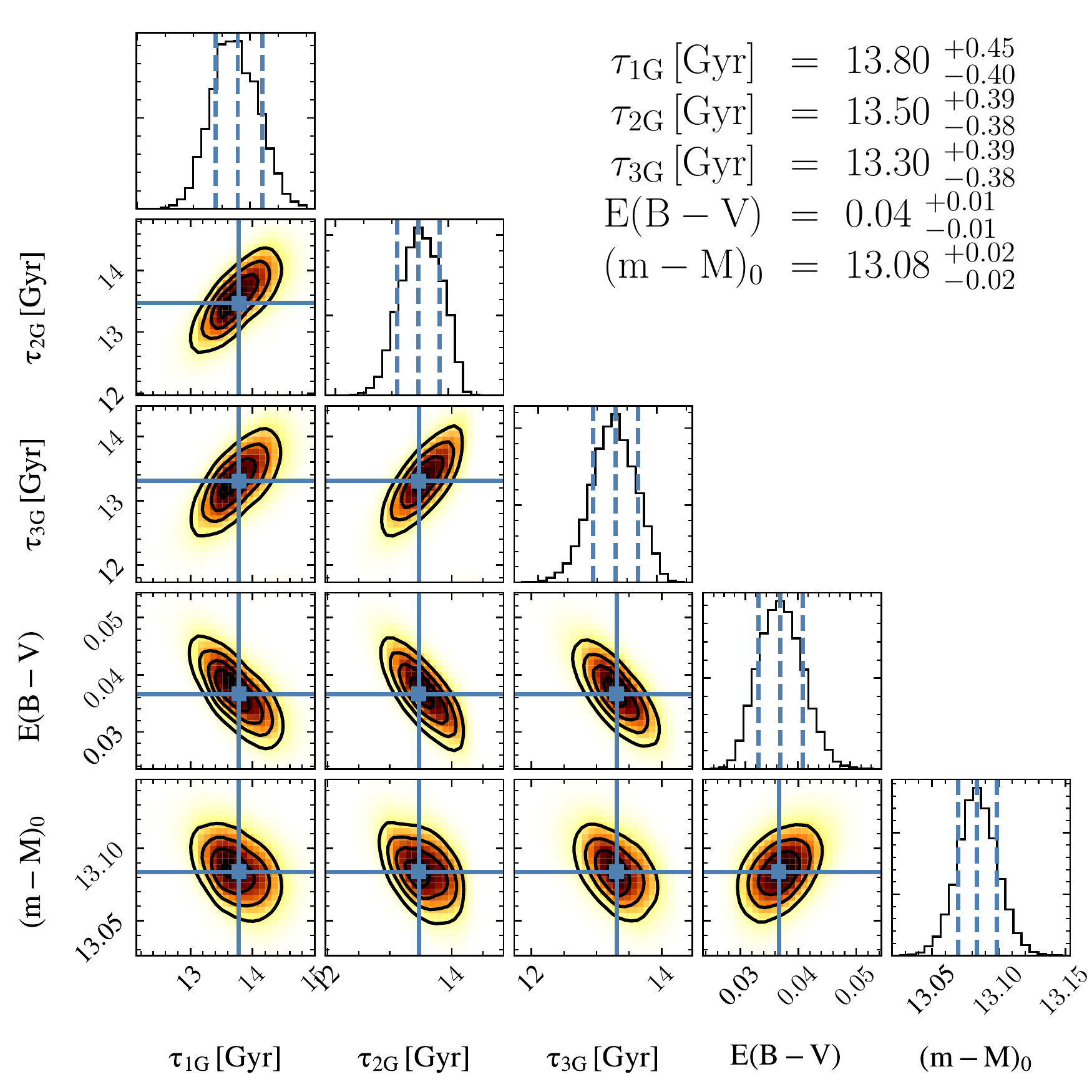}
    \includegraphics[scale=0.5]{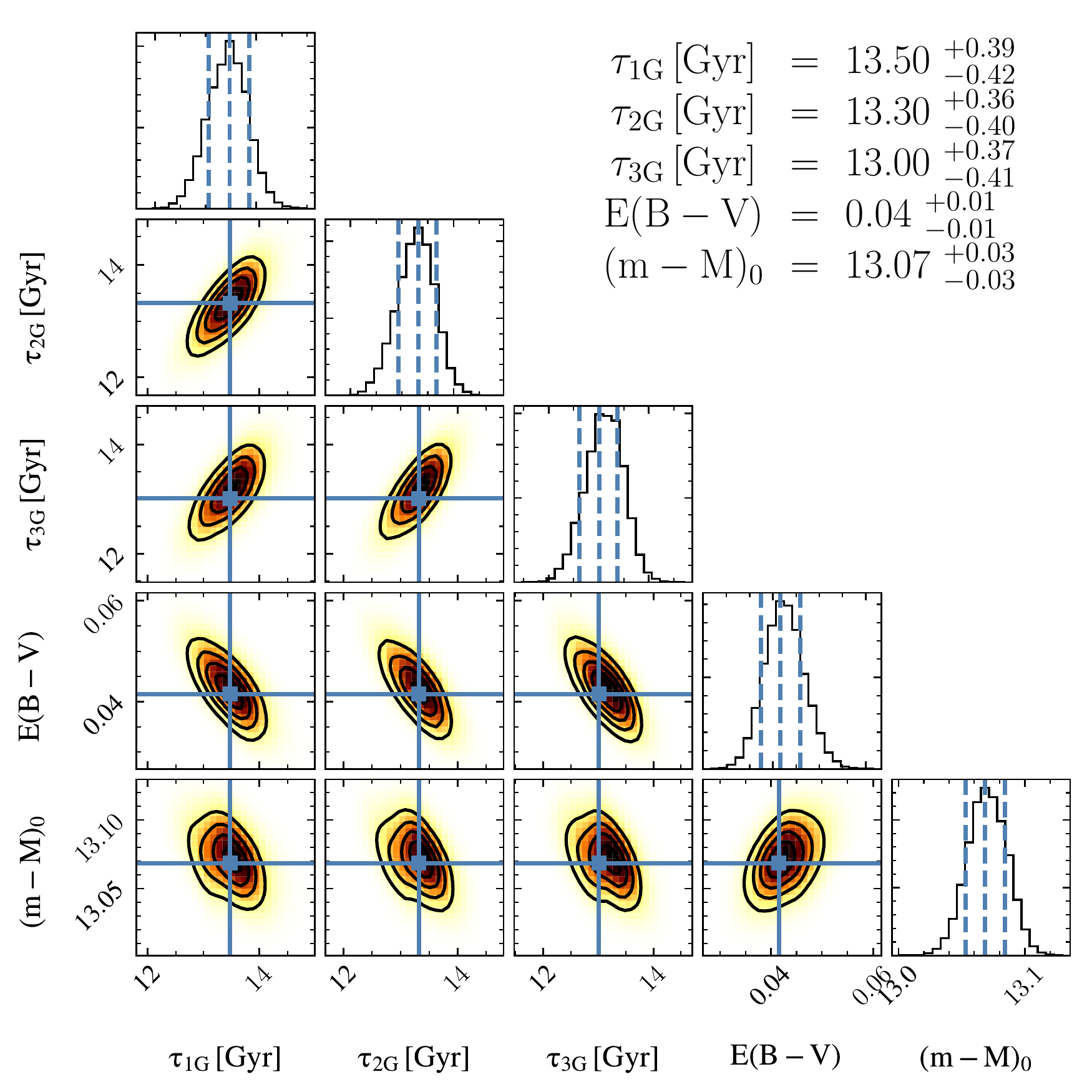}
    \caption{Corner plots for NGC~6752. Left panel: simultaneous fitting of the three stellar populations, adopting canonical helium abundance; Right panel: same as in left panel, but taking into account helium abundance differences.
    }
    \label{fig:ngc6752-dsed_mp}
\end{figure*}

In the following the analysis of NGC 6752 is restricted to DSED isochrones.
The procedure starts with the isochrone fitting assuming 
the CMD to consist of a SSP, and the method is subsequently applied to the MPs. In order to carry out the isochrone fitting, we employed the same CMD m$_{\rm F606W}$ vs. (m$_{\rm F606W}-{\rm m}_{F814W}$) used for the  synthetic-data. In the left panel of Figure~\ref{fig:ngc6752_mps} is shown the CMD of NGC~6752 including all stars
as a SSP. 
The value of [Fe/H] = $-1.48$ dex was used as prior through Gaussian distribution with standard deviation of $0.07$. A prior in distance was applied with the value of apparent distance modulus  $(m-M)_{\rm V} = 13.26\pm0.08$  taken from \citet{gratton03}. The results of SSP isochrone fitting are shown in Table~\ref{tab:ngc6752} and Figures \ref{fig:ngc6752-dsed_ssp} and \ref{fig:ngc6752-dsed_mp}. 
The SSP age derivation of $13.7 \pm 0.5$ Gyr is in good agreement with \citet{gratton03}, that obtained  $13.4\pm1.1$ Gyr, and with the Bayesian technique from \citet{wagnerkaiser17} that resulted in an age of $13.202^{+0.174}_{-0.152}$ Gyr. The parallax from Gaia DR2 \citep{gaia18b} for the NGC~6752, $\bar{\omega} = 0.2610\pm0.0011$ mas, corrected by the zero point of $-0.03$ mas given by \citet{lindegren18}, gives a heliocentric distance of $3.85\pm0.02$ kpc. Considering  NGC~6752 as a SSP, the derived distance is $4.11\pm0.08$ kpc, in agreement within $~3\sigma$ with Gaia DR2.

The metallicity estimated from SSP isochrone fitting, [Fe/H] $ = -1.49^{+0.05}_{-0.05}$, was fixed for the MPs approach. The metallicity can be fixed because no [Fe/H] variation is detected in this cluster.

To derive the age difference between the stellar {populations}, the hierarchical likelihood described in Section \ref{sec:deltatau} with $N=3$ is applied. The fit is carried out simultaneously to 1G, 2G, and 3G. Firstly, we consider the primordial helium content value for all populations. 
In a second run, we assume a helium enhancement by a type of polluter star, changing the amount of helium for each generation, according to values computed by \citet{milone19}: $\delta \rm Y_{\rm 1G,2G} = 0.010$ and $\delta \rm Y_{\rm 1G,3G} = 0.042$ for the 2G, and 3G, respectively (Figures \ref{fig:ngc6752_mps}, \ref{fig:ngc6752-dsed_mp}, and Table \ref{tab:ngc6752}). We assumed the helium enhancement values from Milone et al. (2019) since they  were derived using the same DSED stellar evolutionary models employed here, therefore there is compatibility. For the metallicity of
NGC~6752, the corresponding canonical helium content in the DSED isochrones is $0.247$, which was associated to 1G. The 2G and 3G helium contents were assumed to be of $0.257$ and $0.289$, adopting the $\delta \rm Y$ values from \citet{milone19}.


\input{TAB-Results2.tex}

\begin{figure*}
    \centering
    \includegraphics[scale=0.5]{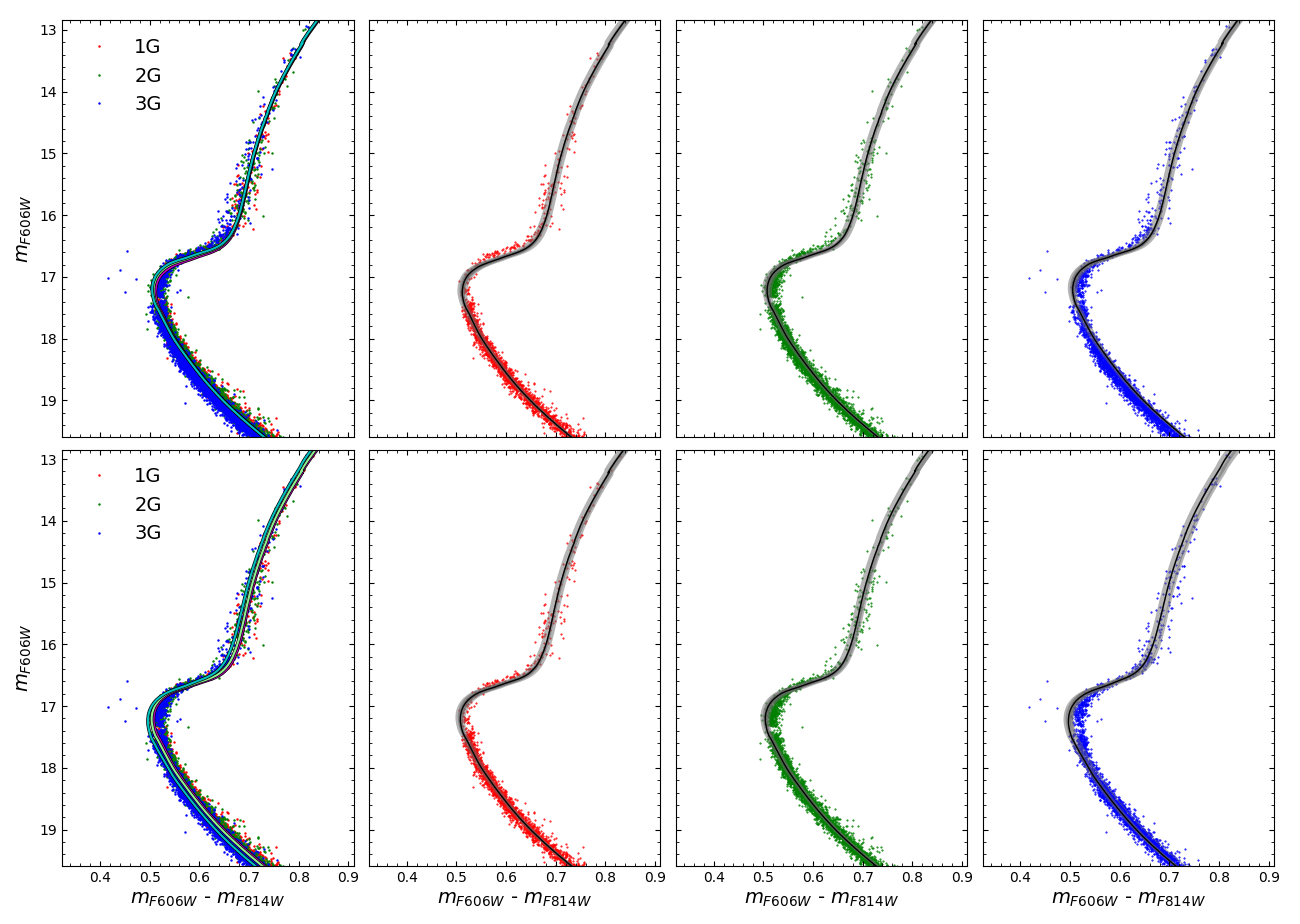}
    \caption{Isochrone fitting for NGC~6752. Left panel: MPs all together.
    Second to fourth panels: isochrone fitting to 1G, 2G, and 3G.
    Upper panels: Canonical helium. Lower panels: Enhanced helium. The strips are the solutions within $1\sigma$.}
    \label{fig:ngc6752-bestfit}
\end{figure*}

Table \ref{tab:ngc6752} and Figure \ref{fig:ngc6752-bestfit} provide the results of isochrone fitting to the MPs. The derived distances using canonical helium and helium enhanced are $4.13\pm0.06$ and $4.11\pm0.08$ kpc, respectively. The latter distance determination is in agreement with the distance from the inverse Gaia DR2 parallax \citep{gaia18b} (see above).
We derive age differences of $\Delta \tau_{\rm 1G,2G} = 300\pm400$ Myr,
and $\Delta \tau_{\rm 1G,3G} = 500\pm400$ Myr, relative to the age of 1G stars, considering that there is no helium enhancement within the GC. However, taking into account the GC helium enhancement cf. \citet{milone19}, 
and noting that the method fits the three {stellar populations} simultaneously,
the 1G is less old (even if its He is still canonical), and
the age differences are of $\Delta \tau_{\rm 1G,2G} = 200\pm400$ Myr,
and $\Delta \tau_{\rm 1G,3G} = 500\pm400$ Myr. These results could give hints on the possible mechanism of GC internal pollution.

{It is interesting to note that, for the He enhanced populations, the result is similar to those with no He enhancement. Assuming the primordial helium for the 1G, 2G, and 3G stars, the $\chi^2$ values are $0.10$, $0.13$, and $0.12$, respectively, resulting in a total value of $0.35$. For He enhanced isochrones, the values of $\chi^2$ are $0.09$, $0.14$, and $0.11$, for the 1G, 2G, and 3G stars, respectively and with a total of $0.34$. Therefore, the fitting using He enhanced isochrones are similarly well-fit}. 

Even though the uncertainties on the age derivation do not take into account the differences
between the stellar evolutionary models, our uncertainty determinations are of the same order of magnitude as those by \citet{monty18}. Given that we did not propagate the uncertainties
from the grid size of the parameter space, the uncertainties given here are the formal errors from MCMC algorithm and they are larger than the ones reported by \citet{wagnerkaiser17}.

\section{Conclusions}\label{sec:conclusion}

 We have developed the \texttt{SIRIUS} code to extract the maximum information from  CMDs of stellar clusters,  through a detailed analysis.
 \texttt{SIRIUS} was tested in terms of synthetic data.
 High precision parameter derivations were obtained with sanity checks  that demonstrate the good performance of the code. Small fluctuations of the solutions were found  in terms of the choice of CMD colors, relative to the input parameters of the synthetic data (Figure \ref{fig:sanity-results}). Applying a Monte Carlo spread of stars, these 
 fluctuations increase somewhat, as can be seen in Table \ref{tab:sanity2}.
 In any case, the solution obtained is within the uncertainties and limited because of the grid resolution in the parameter space.
 
 The \texttt{SIRIUS} code is applied to analyse the halo
 globular cluster NGC 6752 of metallicity [Fe/H]$\approx$-1.49.
 Three stellar populations are identified, confirming previous findings
 by \citet{carretta12} from spectroscopy, and \citet{milone19} from photometry.
 The age derivation of the three stellar {populations}, taking into account
 He abundance differences from \citet{milone19}, results to
 be of $200/300\pm400$ Myr between 1G and 2G and between 2G and 3G.
 This points to a possible interpretation of having the same mechanism
 producing 2G, and later the 3G.
 
 Many authors have extensively
 discussed the probable candidates to produce the chemical abundance
 patterns of second (and subsequent) stellar {populations} from self-enrichment of the cluster. The main
 candidates are the AGB stars, 
 and SMS, in both cases through their winds, as well as FRMSs \citep[][]{decressin07, krause13}. All of them predict an age difference between the stellar populations.
 
 In conclusion, given the uncertainties in the models
 of pollution, and the uncertainties in the age difference
 derived from the CMDs, it is not possible to firmly indicate a
 scenario for the formation of a second {stellar population}. The age differences derived for NGC 6752 could be compatible with the AGB scenario if only the best value determinations are taken into account. However, considering the uncertainties, the results could be compatible with all scenarios regarding the origin of MPs (SMS and FRMS), even those with no age difference. 
 Further analyses of age differences of multiple stellar
 populations are of great interest. In particular, within
 the {\it HST} Legacy survey collaboration, \citet{nardiello15b} derived the relative age of NGC~6352 MPs from $\chi^2$ minimization isochrone fitting, assuming each of them as SSPs, and Oliveira et al. (2019, in preparation) apply the methods described here to derive the ages for seven bulge globular clusters and their MPs.

\acknowledgments
{We acknowledge the anonymous referee for a detailed review and helpful suggestions, which allowed us to improve the manuscript.} SOS acknowledges the FAPESP PhD fellowship 2018/22044-3. LOK and BB acknowledge partial financial support from FAPESP, CNPq, and CAPES - Finance Code 001. APV acknowledges the FAPESP postdoctoral fellowship no. 2017/15893-1. RAPO acknowledges the FAPESP PhD fellowship no. 2018/22181-0. DN acknowledges partial support by the Universit\`a degli Studi
di Padova Progetto di Ateneo BIRD178590. APV and SOS acknowledge the DGAPA-PAPIIT grant IG100319.



\bibliography{sirius_ngc6752}{}
\bibliographystyle{aasjournal}



\end{document}

%% file: TAB-MockParams.tex
\begin{table}
    \centering
    \caption{Input parameters for the construction of the synthetic catalogues.}
    \begin{tabular}{l l l}
    \hline
    \noalign{\smallskip}
    \hline
    \noalign{\smallskip}
    { Parameter}                          & No-Spread      & Spread\\
    \noalign{\smallskip}
    \hline
    \noalign{\smallskip}
    Evolutionary Model                     & DSED          & DSED \\
    N$_{\rm total}$                        & $260$         & $10,000$\\
    $\tau_{\rm SSP}$\,(Gyr)                & $13.0$        & $13.0$  \\
    $\Delta \tau$\,(Gyr)                   & --            & $0.1,\,0.5,\,1.5$ \\ 
    $\rm [Fe/H]$\,(dex)                    & $-1.26$       & $-1.26$\\
    $E(B-V)$                               & $0.18$        & $0.18$\\
    (m-M)$_0$                              & $14.38$       & $14.38$\\
    f$_{\rm bin}$                          & --            & $0.30$ \\
    q$_{\rm min}$                          & --            & $0.60$ \\
    ${\rm N}_{\rm 1G}/{\rm N}_{\rm total}$ & $1.000$       & $0.360$\\
    \hline
    \hline
    \end{tabular}
    \label{tab:my_label}
\end{table}

%% file: TAB-ResSyn.tex
\begin{table*}
    \centering
    \caption{Sanity check with spread data, results summarized for synthetic-data in SSP context and MPs.}
    \begin{tabular}{c c l c c c c c }
    \hline
    \hline
    \noalign{\smallskip}
    \multirow{2}{*}{Sanity Check} & \multirow{2}{*}{N$_{\rm 1G}$/N$_{\rm Tot}$} & \multirow{2}{*}{Model} & $\tau_{\rm SSP}$ & $\Delta \tau_{\rm 1G,2G}$ & [Fe/H] & $E(B-V)$ & $(m-M)_0$ \\
    \noalign{\smallskip}
      &   &   & (Gyr) & (Gyr) & (dex) & (mag) & (mag) \\
    \noalign{\smallskip}
    \hline
    \noalign{\smallskip}
    \multirow{2}{*}{ SSP} & \multirow{2}{*}{ -- } & DSED  & $12.70^{+0.36}_{-0.37}$ & -- & $-1.26^{+0.03}_{-0.03}$ & $0.18^{+0.01}_{-0.01}$ & $14.35^{+0.03}_{-0.03}$\\
    \noalign{\smallskip}
                             &   & BaSTI & $13.80^{+0.61}_{-0.61}$ & -- & $-1.26^{+0.03}_{-0.03}$ & $0.18^{+0.01}_{-0.01}$ & $14.30^{+0.04}_{-0.03}$\\
    \noalign{\smallskip}
    \hline
    \noalign{\smallskip}
    \multirow{2}{*}{  MPs $\Delta \tau=0.10$ Gyr} & \multirow{2}{*}{ $0.377\pm0.011$ } & DSED  &  -- & $0.11^{+0.36}_{-0.38}$ & $-1.26^{+0.02}_{-0.03}$ & $0.18^{+0.01}_{-0.01}$ & $14.38^{+0.03}_{-0.03}$\\
    \noalign{\smallskip}
                            &    & BaSTI & -- & $0.19^{+0.49}_{-0.49}$ & $-1.26^{+0.03}_{-0.03}$ & $0.18^{+0.01}_{-0.01}$ & $14.33^{+0.03}_{-0.03}$\\
    \noalign{\smallskip}
    \hline
    \noalign{\smallskip}
    \multirow{2}{*}{ MPs $\Delta \tau=0.50$ Gyr} & \multirow{2}{*}{ $0.370\pm0.012$ } & DSED  &  -- & $0.41^{+0.43}_{-0.37}$ & $-1.26^{+0.03}_{-0.02}$ & $0.18^{+0.01}_{-0.01}$ & $14.38^{+0.03}_{-0.03}$\\
    \noalign{\smallskip}
                            &    & BaSTI & -- & $0.51^{+0.54}_{-0.54}$ & $-1.26^{+0.02}_{-0.02}$ & $0.18^{+0.01}_{-0.01}$ & $14.33^{+0.03}_{-0.03}$\\
    \noalign{\smallskip}
    \hline
    \noalign{\smallskip}
    \multirow{2}{*}{ MPs $\Delta \tau=1.50$ Gyr} & \multirow{2}{*}{ $0.339\pm0.008$ } & DSED  & -- & $1.20^{+0.44}_{-0.38}$ & $-1.26^{+0.02}_{-0.03}$ & $0.18^{+0.01}_{-0.01}$ & $14.37^{+0.03}_{-0.03}$\\
    \noalign{\smallskip}
                            &    & BaSTI & -- & $1.47^{+0.53}_{-0.46}$ & $-1.26^{+0.03}_{-0.02}$ & $0.18^{+0.01}_{-0.01}$ & $14.35^{+0.03}_{-0.03}$\\
    \noalign{\smallskip}
    \hline
    \hline
    
    \end{tabular}
    \label{tab:sanity2}
\end{table*}

%% file: TAB-Results2.tex
\begin{table*}
    \centering
    \caption{Results of isochrone fitting for NGC~6752 in SSP context and MPs.}
    \begin{tabular}{ c c c c c c c c c c  c }
    \hline
    \hline
    \noalign{\smallskip}
      & Y  & $\tau$ & $\Delta \tau_{\rm 1G,2G}$ & $\Delta \tau_{2G,3G}$  & $\Delta \tau_{1G,3G}$ & [Fe/H] & $E(B-V)$ & $(m-M)_0$ & $(m-M)_{\rm V}$ & d$_{\odot}$\\
    \noalign{\smallskip}
     &  & (Gyr) & (Gyr) & (Gyr)  & (Gyr) & (dex) & & & & (kpc)\\ 
    \noalign{\smallskip}
    \hline
    \noalign{\smallskip}
     SSP & Y(Z)$^{\dagger}$  & $13.70^{+0.61}_{-0.38}$ & --  & -- & -- & $-1.49^{+0.05}_{-0.05}$ & $0.03^{+0.02}_{-0.02}$ & $13.07^{+0.03}_{-0.03}$ & $13.16^{+0.07}_{-0.07}$ & $4.11\pm0.08$  \\
    \noalign{\smallskip}
    \hline
    \noalign{\smallskip}
    \multicolumn{11}{c}{MPs with Y canonical} \\
    \noalign{\smallskip}
    \hline
    \noalign{\smallskip}
      1G & $0.247$   & $13.80^{+0.45}_{-0.40}$ & \multirow{4}{*}{$0.30^{+0.42}_{-0.39}$} & \multirow{4}{*}{$0.20^{+0.39}_{-0.38}$} & \multirow{4}{*}{$0.50^{+0.43}_{-0.39}$} & \multirow{4}{*}{$-1.49^{\dagger\dagger}$} & \multirow{4}{*}{$0.04^{+0.01}_{-0.01}$} & \multirow{4}{*}{$13.08^{+0.02}_{-0.02}$} & \multirow{4}{*}{$13.20^{+0.03}_{-0.03}$} & \multirow{4}{*}{$4.13\pm0.06$}  \\
    \noalign{\smallskip}
     2G & $0.247$  & $13.50^{+0.39}_{-0.38}$ & & &  & & &  & &\\
    \noalign{\smallskip}
     3G & $0.247$ & $13.30^{+0.39}_{-0.38}$ & & &  & & &  & &\\
    \noalign{\smallskip}
    \hline
    \noalign{\smallskip}
    \multicolumn{11}{c}{MPs with Y enhancement}\\
    \noalign{\smallskip}
    \hline
    \noalign{\smallskip}
     1G  & $0.247$ & $13.50^{+0.39}_{-0.42}$ & \multirow{4}{*}{$0.20^{+0.38}_{-0.41}$} & \multirow{4}{*}{$0.30^{+0.37}_{-0.41}$} & \multirow{4}{*}{$0.50^{+0.38}_{-0.42}$} & \multirow{4}{*}{$-1.49^{\dagger\dagger}$} & \multirow{4}{*}{$0.04^{+0.01}_{-0.01}$} & \multirow{4}{*}{$13.07^{+0.03}_{-0.03}$} & \multirow{4}{*}{$13.19^{+0.03}_{-0.03}$} & \multirow{4}{*}{$4.11\pm0.08$} \\
    \noalign{\smallskip}
     2G & $0.257$ & $13.20^{+0.39}_{-0.41}$ & & &  & & &  & & \\
    \noalign{\smallskip}
     3G & $0.289$ & $13.00^{+0.41}_{-0.41}$ & & &  & & &  & & \\
    \noalign{\smallskip}
    \hline  
    \hline
    \noalign{\smallskip}
    \multicolumn{11}{l}{$\dagger$ Y as function of Z, defined by: $0.245 + 1.5 \times \rm Z$.}\\
    \multicolumn{11}{l}{$\dagger\dagger$ Fixed value from the SSP isochrone fitting.}
    \end{tabular}
    \label{tab:ngc6752}
\end{table*}